\documentclass[aps,prd,reprint,nofootinbib,superscriptaddress]{revtex4-2}
\usepackage{amsmath,amssymb,mathtools}
\usepackage{mathrsfs}
\usepackage{bm}
\usepackage{graphicx}
\usepackage[colorlinks=true,linkcolor=blue,citecolor=blue,urlcolor=blue]{hyperref}
\usepackage{enumitem}
\usepackage{xcolor}
\usepackage{comment}

\begin{document}
\preprint{CHIBA-EP-283}
\title{
Gauge-Invariant Brout--Englert--Higgs Mechanism \\
without Spontaneous Breaking of Local Gauge Symmetry: \\
A Quantum-Reference-Frame Formulation
}

\author{Kei-Ichi Kondo}
\email{kondok@faculty.chiba-u.jp}
\affiliation{Department of Physics, Graduate School of Science, Chiba University, 1-33 Yayoi-cho, Chiba, Chiba 263-8522, Japan}
\affiliation{Research and Education Center for Natural Sciences, Keio University, 4-1-1 Hiyoshi, Yokohama, Kanagawa 223-8521, Japan}
\date{October 11, 2026}

\begin{abstract}

The Brout--Englert--Higgs (BEH) mechanism can be formulated in gauge-invariant relational terms without assigning physical spontaneous breaking to a local gauge redundancy. 
We treat the gauge orientation of the Higgs field as dynamical internal reference data and describe the gauge connection relative to this Higgs frame. 
On a regular patch where the frame exists, the Higgs kinetic term induces a positive-semidefinite mass-squared quadratic form on the gauge-orbit directions and reproduces the standard tree-level vector masses. 
A complete Higgs frame yields a gauge-invariant relational connection, whereas a partial frame is defined only up to a stabilizer group $H$ and produces Higgs stiffness only along the coset directions $G/H$. 
In quantum field theory the normalized frame and relational connection are nonpolynomial composite variables; 
physical masses are therefore characterized through suitably renormalized gauge-invariant composite correlators, or through fully dressed operators in residual-charged sectors. 
Applying this construction to the electroweak theory, we show that the Higgs frame relationalizes the $SU(2)_L$ redundancy while leaving a residual electromagnetic $U(1)_{\rm em}$ Gauss law: $W^\pm$ and charged fermionic modes require electromagnetic dressing, whereas neutral channels admit local gauge-invariant interpolators. 
This formulation connects Higgs dressings, FMS composites, and internal quantum-reference-frame language while keeping local mass generation distinct from global symmetry realization and topological obstructions.

\end{abstract}

\maketitle

\tableofcontents

\section{Introduction}
\label{sec:intro}

The conventional Brout--Englert--Higgs (BEH) description for gauge-boson mass generation is commonly presented in terms of spontaneous symmetry breaking and gauge-fixed elementary fields \cite{EnglertBrout1964,Higgs1964,GuralnikHagenKibble1964,
Higgs1966,Kibble1967}. 
Gauge-invariant accounts of the Higgs phenomenon, in which physical local-gauge symmetry breaking is not taken as the defining mechanism, have a substantial history, including the Froehlich--Morchio--Strocchi (FMS) construction \cite{FMS1980,FMS1981} and later gauge-invariant analyses \cite{Struyve2011,Kondo2016PLB,Kondo2017EPJ,Kondo2018}.
See e.g. \cite{Maas2019} for a review.

\textbf{The present paper develops a specifically Quantum Reference Frame (QRF)-based organization of the BEH mechanism.} 
\textbf{The orientation of the Higgs field is treated as a dynamical internal reference frame, and the gauge connection is described relationally with respect to that frame.}
\textbf{The Higgs kinetic term then measures the energetic cost of relative motion between the connection and the frame and reproduces the usual tree-level BEH mass matrix without using a gauge-variant Higgs expectation value as the physical criterion.}

From this viewpoint, the usual explanation that ``the vacuum expectation value chooses an orientation and breaks gauge symmetry'' is a gauge-fixed representation of the same dynamics rather than the gauge-invariant definition of the mechanism. 
In fact, Elitzur's theorem makes clear that, without gauge fixing, a local gauge-variant field cannot serve as an ordinary spontaneous order parameter \cite{Elitzur1975}.

\textbf{
The aim of this paper is therefore not to claim priority for using Higgs matter as reference data, but to systematize complete and partial Higgs frames, gauge-invariant massive modes, FMS composites, electroweak residual gauge structure and quantum/nonperturbative relational diagnostics 
within one internal-QRF framework.
}

The main contributions of the present formulation are:

\noindent
(i) a common QRF dictionary for complete Higgs frames and partial $G/H$ frames;

\noindent
(ii) a direct identification of the Higgs kinetic stiffness with the classical BEH mass matrix;

\noindent
(iii) a separation between the exact dressed-variable identity underlying the FMS vector composite and the perturbative FMS pole matching \cite{FMS1980,FMS1981};

\noindent
(iv) an explicit relation between Kondo's gauge-covariant/gauge-invariant massive modes \cite{Kondo2016PLB,Kondo2018} and gauge-invariant QRF components;

\noindent
(v) an electroweak treatment retaining the residual $U(1)_{\rm em}$ Gauss law; and

\noindent
(vi) a quantum and nonperturbative research program involving fluctuating frames, relational effective actions, defects, and confinement diagnostics.


The logical structure of the paper can be summarized by three structural statements.  

\textbf{Complete-frame statement:} on a regular complete-frame patch, $(\mathscr A_\mu,\Phi)/G$ is locally represented by relational variables $(\mathscr A_\mu^h,\Phi^h)$, and the Higgs kinetic term generates the corresponding mass-squared quadratic form.  

\textbf{Partial-frame statement:} when the Higgs configuration has stabilizer (stability subgroup) $H$, a local lift is defined only up to $H$; the relational connection separates into a residual $H$ connection and $G/H$ coset components, and the kernel of the Higgs-generated mass matrix is the stabilizer algebra.  

\textbf{Electroweak corollary:} after the Higgs frame relationalizes the $SU(2)_L$ redundancy, the residual local action on relational fields is generated by $Q=T^3+Y$, so charged $W^\pm$ and charged fermionic modes require an electromagnetic dressing for fully physical charged states.

This paper is organized as follows. Section~\ref{sec:gauge-redundancy} explains why physical spontaneous breaking of local gauge redundancy is not required for the BEH mechanism, and Sec.~\ref{sec:qrf-dictionary} introduces the basic relational/QRF dictionary. Section~\ref{sec:abelian} develops the Abelian prototype, while 
Secs.~\ref{sec:nonabelian} and \ref{sec:partial} treat complete and partial non-Abelian Higgs frames.
Section~\ref{sec:fms} connects the relational variables with the FMS description, and 
Sec.~\ref{sec:electroweak} applies the construction to the electroweak theory. 
Section~\ref{sec:remarks} discusses 
Gauss's law, 
the would-be Nambu--Goldstone sector, 
defects, 
unitary gauge, 
the equivalence theorem,  
the Stueckelberg limit, and 
the dressing-field method. 
Section~\ref{sec:definition} summarizes the operational definition and dictionary of QRF-BEH. 
Conclusions and open problems are given in Sec.~\ref{sec:conclusion}.


Appendix~\ref{app:adj-Higgs} gives the details of the derivation of the relational gauge field in the adjoint--Higgs model.

Appendices~\ref{app:ew-derivation} and \ref{app:jacobian} provide the electroweak kinetic-term derivation and the local Jacobian/notation conventions.

\section{Why physical spontaneous breaking of local gauge redundancy is not required}
\label{sec:gauge-redundancy}

\subsection{The conventional BEH mechanism based on spontaneous breaking of gauge symmetry}

The standard textbook description starts from a decomposition of the complex scalar field in the simplest Abelian Higgs model such as
\begin{equation}
 \phi(x)=\frac{\rho(x)}{\sqrt{2}}e^{i\vartheta(x)}
 =\frac{v+H(x)}{\sqrt{2}}e^{i\vartheta(x)} \in \mathbb{C},
\end{equation}
and says that a non-vanishing vacuum expectation value $\langle \phi \rangle\not=0$ spontaneously breaks the gauge symmetry $\phi(x)\mapsto e^{i\alpha(x)}\phi(x)$ and that the Nambu--Goldstone mode associated with the spontaneous symmetry breaking is eaten by the gauge boson.  
This language is convenient in gauge-fixed perturbation theory, but it is not the most direct statement of the gauge-invariant physics.

A local gauge transformation is a redundancy in the description.
Elitzur's theorem \cite{Elitzur1975} shows that, 
\textbf{in a local gauge theory without gauge fixing, the expectation value of a gauge-variant local order parameter cannot serve as an ordinary physical order parameter for spontaneous breaking of a local gauge symmetry}. Therefore, \textbf{the non-vanishing vacuum expectation value (VEV)
\begin{equation}
 \langle \Phi\rangle\neq 0
\end{equation}
should not be regarded as the physical definition of the BEH mechanism.}

What can have direct physical meaning are gauge-invariant observables, genuine global symmetries, and spectra extracted from gauge-invariant operators.  The FMS construction makes this especially explicit: physical scalar and vector excitations can be represented by gauge-invariant composite operators, while ordinary perturbative particle poles are recovered in the appropriate expansion \cite{FMS1980,FMS1981,Maas2019}.

\subsection{Local gauge symmetry does not imply physical vacuum degeneracy}

Consider the Mexican-hat potential $V(\phi^*\phi)$ of a complex Higgs field $\phi$. Ordinarily, the circumference of the bottom, or its higher-dimensional sphere, is called a set of ``
\textbf{degenerate vacua}
.'' However, if that direction merely lies along a local gauge orbit, these are not distinct physical states.

In the simplest Abelian $U(1)$ Higgs model with one complex Higgs field $\phi$, 
pointwise in the Higgs fiber, quotienting a nonzero complex Higgs field $\phi$ by the local $U(1)$ phase $\vartheta$ leaves only the half-line $\rho\geq0$; the angular circle $\vartheta$ is one gauge orbit. 
The local Higgs potential is a potential $V(\rho^2/2)$ for the gauge-invariant radial variable $\rho$, with its radial minimum at $\rho=v$.

Likewise, in the non-Abelian, e.g., $SU(2)$ Higgs model with one fundamental Higgs field $\Phi$, the angular part of a nonzero Higgs configuration is $S^3\simeq SU(2)$, which is precisely the local $SU(2)$ gauge orbit. 
In the case of complete Higgsing, the physical radial configuration is again essentially $\rho\geq0$, and there is no physical vacuum degeneracy originating from the local gauge orbit.

\subsection{What can be broken is a genuine global symmetry}

Local gauge redundancy should be distinguished from custodial symmetry, flavor symmetry, and global symmetries that act nontrivially at a boundary or at infinity. 
Genuine global, custodial, and asymptotic symmetry realizations may exhibit state-dependent symmetry breaking, but different gauge-fixing remnant symmetries need not define the same physical transition \cite{CaudyGreensite2008}. Such symmetry breaking is logically distinct from, and not required by, the existence of a classical Higgs-induced vector mass term.

Accordingly, in the QRF-BEH formulation the following two are strictly separated:
\begin{enumerate}[label=(\arabic*)]
 \item BEH mass generation: relational dynamics of the Higgs QRF and the gauge field.
 \item Genuine global symmetry breaking: state-dependent realization of custodial, flavor, or asymptotic symmetries.
\end{enumerate}
The former can exist without the latter.

\subsection{Reformulation of the BEH mechanism from QRF}

The BEH mechanism may therefore be reformulated as follows.
\textbf{
The Higgs field supplies an internal reference frame.
}
The Higgs kinetic term generates a positive-semidefinite quadratic form in the internal gauge-orbit directions.  In the Euclidean action, or equivalently in the Hamiltonian energy of the physical modes, this can be interpreted as \textbf{an energy cost for a mismatch between parallel transport by the gauge connection and the local orientation of the Higgs frame}.  
The resulting relational vector has the standard BEH mass at tree level.
In compact form,
\begin{equation}
 \begin{aligned}
 \text{BEH mechanism}
 ={}&\text{dynamical internal QRF}\\
 &+\text{relative-connection stiffness}\\
 &+\text{massive relational vector}.
 \end{aligned}
\end{equation}
No physical spontaneous breaking of the local gauge redundancy is required in this statement.

In what follows we exemplify this viewpoint from QRF for the Abelian and non-Abelian gauge--Higgs models with the complete and partial BEH mechanism.

\section{Basic QRF dictionary}
\label{sec:qrf-dictionary}

\subsection{Quantum reference frame and relational fields}

Let $G$ be the local gauge group.
The \textbf{matter field} $\Phi(x)$ in the fundamental representation has the local gauge transformation
\begin{equation}
 \Phi(x)\mapsto \Omega(x)\Phi(x), \quad \Omega(x)\in G .
\end{equation}
The non-Abelian gauge field denoted by
\begin{equation}
 \mathscr{A}_\mu(x)=\mathscr{A}_\mu^A(x) T_A  ,
\end{equation}
has the local gauge transformation:
\begin{align}
 \mathscr{A}_\mu(x) \mapsto{}&
 \Omega(x)\mathscr{A}_\mu(x)\Omega(x)^{-1}
 \nonumber\\
 &+\frac{i}{g}\Omega(x)\partial_\mu\Omega(x)^{-1},
 \qquad \Omega(x)\in G.
\end{align}

Let $h(x)$ be a group-valued \textbf{frame field $h[\Phi](x)$ constructed from matter fields} $\Phi$. When 
\textbf{a complete internal QRF} 
exists, the frame field $h(x)$ transforms under a local gauge transformation $\Omega(x)\in G$ from the left as
\begin{equation}
 h(x)\mapsto \Omega(x)h(x), \quad \Omega(x)\in G .
\end{equation}
The relational matter field $\Phi^h(x)$ relative to the Higgs QRF $h(x)$ is introduced by
\begin{equation}
\boxed{
  \Phi^h(x):=h(x)^{-1}\Phi(x),
}
\end{equation}
so that  $\Phi^h(x)$ is invariant under the original local gauge transformation acting from the left:
\begin{equation}
 \Phi^h(x)\mapsto \Phi^h(x).
\end{equation}

The \textbf{relational non-Abelian gauge field} or the \textbf{relational connection} $\mathscr{A}_\mu^h(x)$ relative to the Higgs QRF $h$ is introduced by \cite{Donnelly:2016auv,GomesRiello:2016,Gomes:2018shn,CarrozzaHoehn2022}
\begin{equation}
 \boxed{
 \mathscr{A}_\mu^h(x) := h(x)^{-1}\mathscr{A}_\mu(x) h(x)+\frac{i}{g}h(x)^{-1}\partial_\mu h(x)
 }
 \label{eq:relconnection}
\end{equation}
Then, 
$\mathscr{A}_\mu^h(x)$ is invariant under the original local gauge transformation acting from the left:
\begin{equation}
 \mathscr{A}_\mu^h(x)\mapsto \mathscr{A}_\mu^h(x).
\end{equation}
$\mathscr{A}_\mu^h(x)$ is the gauge connection measured relative to the dynamical matter frame $h[\Phi]$, not relative to an externally supplied color basis.  The dressed/relational variable is defined without imposing a gauge condition on the functional integral.  On a regular patch, however, the same field-dependent dressing is closely associated with the representative in which the Higgs orientation is fixed, namely the Higgs-adapted or unitary-gauge representative.  Thus the invariant relational variable should not be confused with a gauge-fixed bare field, even though their component representatives coincide when $h=1$.

In the classical configuration space, $h[\Phi]$ and $\mathscr A_\mu^h$ are exact relational variables on a regular patch.  In continuum quantum field theory, however, a normalized frame such as $h[\Phi]$ and the resulting $\mathscr A_\mu^h$ are nonpolynomial composite quantities and need not define renormalized pointlike operators without an additional prescription.  Quantum spectral statements below are therefore formulated primarily in terms of suitably renormalized polynomial gauge-invariant composites whenever possible; $h$ and $\mathscr A_\mu^h$ should otherwise be regarded as relational patch variables.

More precisely, on a regular complete-frame patch the Higgs frame $h[\Phi](x)$ is an equivariant map,
\begin{equation}
 h[\Omega\Phi](x)=\Omega(x)h[\Phi](x).
 \label{eq:complete-equivariance}
\end{equation}
For a partial frame with stabilizer (stability subgroup) $H$, a local lift obeys the corresponding equivariance law only up to a compensator,
\begin{equation}
 h[\Omega\Phi](x)=\Omega(x)h[\Phi](x)r(x;\Omega,\Phi)^{-1},
 \qquad r(x;\Omega,\Phi)\in H.
 \label{eq:partial-equivariance-preview}
\end{equation}
These equivariance properties distinguish the reference data from an externally fixed internal basis and provide the starting point for changes between admissible internal QRFs.

Similarly, the corresponding relational field strength is given by
\begin{equation}
 \mathscr{F}_{\mu\nu}^{h}(x):=h(x)^{-1}\mathscr{F}_{\mu\nu}(x)h(x),
\end{equation}
which is also locally gauge invariant:
\begin{equation}
 \mathscr{F}_{\mu\nu}^{h}(x)\mapsto\mathscr{F}_{\mu\nu}^{h}(x).
\end{equation}

The physical principle is that the internal orientation of either the system or the frame separately is gauge dependent, whereas their relative orientation is gauge invariant.
This mathematical structure is closely related to the dressing-field method, in which a dressing constructed from the field content is used to form gauge-invariant composite variables \cite{FournelEtAl2012,BerghoferFrancois2024}.  In QRF language the additional physical emphasis is that the dressing is itself a dynamical reference subsystem and fluctuates in quantum theory.

The \textbf{basic principle of QRF} is that \textbf{the physical quantity obeys not the ``bare internal orientation'' but the ``relative orientation between system and frame.''}
In studies of field-space connections as well, \textbf{the idea of using the matter field itself as a reference frame to separate gauge and physical directions has been made explicit}, and a relation between the Higgs connection and quantum reference frames has been pointed out \cite{Gomes:2018shn}.

\subsection{Admissible frame changes and frame-independent physical masses}

Let $h$ and $h'$ be two admissible complete Higgs frames on the same regular patch $h'=hs$.  Since both transform equivariantly from the left, their relative orientation
\begin{equation}
 s(x):=h(x)^{-1}h'(x)
 \label{eq:relative-frame-s}
\end{equation}
is invariant under the original local gauge redundancy.  The corresponding relational connections are related by
\begin{equation}
 \mathscr A_\mu^{h'}
 =s^{-1}\mathscr A_\mu^h s+\frac{i}{g}s^{-1}\partial_\mu s.
 \label{eq:qrf-frame-change}
\end{equation}
Equation~\eqref{eq:qrf-frame-change} is a change of internal reference description between two relational frames, not a transformation by the original gauge redundancy.  For a partial frame, the special change of lift $h\mapsto hk$, $k\in H$, is precisely the residual-$H$ transformation of the relational variables.

The Higgs frame can make the BEH quadratic term particularly local and transparent, but the frame-independence statement is restricted to regular, invertible and admissible local or quasilocal changes of relational variables within the same physical and asymptotic sector.  At tree level, for a partial frame the change $h\mapsto hk$ acts on the coset components by the adjoint action of $H$; because the scalar kinetic metric is $H$ invariant, the coset mass quadratic form and hence its eigenvalues are unchanged.  For complete frames, an admissible change such as Eq.~\eqref{eq:qrf-frame-change} changes the interpolating variables but not the spectrum of the underlying gauge-invariant theory.  Singular, noninvertible, or asymptotically inequivalent changes are not covered by this statement.

\section{Abelian $U(1)$ Higgs model: a complete prototype of QRF-BEH}
\label{sec:abelian}

\subsection{Change of variables in a smooth complete-frame patch}

The Lagrangian of the Abelian $U(1)$ gauge--Higgs model is written in terms of the complex scalar field $\phi$ and the $U(1)$ gauge field $A_\mu$ as
\begin{align}
 &\mathscr{L}
 =-\frac14 F_{\mu\nu}[A]F^{\mu\nu}[A]
 +(D_\mu[A]\phi)^*D^\mu[A]\phi
 -V(\phi^*\phi), \nonumber\\
 &F_{\mu\nu}[A] := \partial_\mu A_\nu - \partial_\nu A_\mu ,
 \quad D_\mu[A]:=\partial_\mu-iq A_\mu.
\end{align}
Let the gauge transformation for $\phi$ and $A_\mu$ be
\begin{equation}
 \phi\to e^{iq\alpha}\phi,
 \qquad
 A_\mu\to A_\mu+\partial_\mu\alpha .
\end{equation}

In this prototype $q$ denotes the coefficient multiplying $A_\mu$ in the covariant derivative and no residual group $\mathbb Z_q$ is implied by this notation.  If one wishes to discuss the global compact-$U(1)$ structure, it is useful to separate a minimal charge unit $e$ from an integer Higgs charge $N$, so that $D_\mu\phi=(\partial_\mu-iNeA_\mu)\phi$.  For a $2\pi$-periodic $U(1)$ parameter $\alpha$, the stabilizer of the Higgs phase is then $\mathbb Z_N$: $N=1$ gives a complete phase frame, whereas $N>1$ supplies a frame for $U(1)/\mathbb Z_N$.

Write the complex scalar field $\phi(x)$ in the polar decomposition form:
\begin{equation}
\phi(x)=\frac{\rho(x)}{\sqrt2}e^{i\vartheta(x)} \in \mathbb{C} , \quad \rho \in [0, \infty), \quad  \vartheta\in [0, 2\pi].
\label{polar}
\end{equation}
Then the gauge transformation for ($\rho, \vartheta, A_\mu$) reads
\begin{equation}
\rho \mapsto \rho, \quad \vartheta\mapsto\vartheta+q\alpha,
\qquad
 A_\mu\mapsto A_\mu+\partial_\mu\alpha .
\end{equation}
By substituting (\ref{polar}) into the kinetic term, it is written in terms of ($\rho, \vartheta, A_\mu$) reads
\begin{equation}
|D_\mu[A]\phi|^2
=
\frac12(\partial_\mu\rho)^2
+\frac12\rho^2
(\partial_\mu\vartheta-qA_\mu)
(\partial^\mu\vartheta-qA^\mu).
\end{equation}
We can introduce the new field $W_\mu$ defined by
\begin{equation}
 \boxed{
 W_\mu
 :=A_\mu-\frac1q\partial_\mu\vartheta
 } ,
\end{equation}
which is completely gauge invariant
\begin{equation}
  W_\mu
 \mapsto
 A_\mu+\partial_\mu\alpha
 -\frac1q\partial_\mu(\vartheta+q\alpha)=W_\mu.
\end{equation}
Thus the scalar kinetic term written originally in terms of ($\phi, A_\mu$) or ($\rho, \vartheta, A_\mu$) is identically rewritten in terms of gauge-invariant variables ($\rho, W_\mu$) as
\begin{equation}
\boxed{
 |D_\mu[A]\phi|^2
 =\frac12(\partial_\mu\rho)^2
 +\frac12 q^2\rho^2 W_\mu W^\mu .
}
 \label{eq:abeliankinetic}
\end{equation}

In any smooth complete-frame patch with $\rho\neq0$ in which $\vartheta$ is single-valued, the field strengths agree:
\begin{align}
  F_{\mu\nu}[W] = \partial_\mu W_\nu - \partial_\nu W_\mu
 = \partial_\mu A_\nu - \partial_\nu A_\mu =F_{\mu\nu}[A].
\end{align}
This equality is local to such a smooth patch. In vortex or other nontrivial topological defect sectors, $\vartheta$ can require multiple patches and the distributional commutator $[\partial_\mu,\partial_\nu]\vartheta$ can carry defect flux. The global theory must then be described with the corresponding patching data.
Hence the Lagrangian can be rewritten in terms of gauge-invariant variables ($\rho, W_\mu$) as
\begin{align}
 \mathscr L={}&-\frac14F_{\mu\nu}[W]F^{\mu\nu}[W]
 +\frac12(\partial_\mu\rho)^2
 \nonumber\\
 &+\frac12q^2\rho^2W_\mu W^\mu
 -V\left(\frac12\rho^2\right).
\end{align}
This is a change of the original field variables from ($\phi, A_\mu$) or ($\rho, \vartheta, A_\mu$) to gauge-invariant variables ($\rho, W_\mu$), not a gauge choice.

The same invariant vector also has the parameterization-independent operator form emphasized in the gauge-independent Abelian Higgs formulation of Ref.~\cite{Kondo2017EPJ}. 
Below, $D_\mu^{(+q)}=\partial_\mu-iq A_\mu$ acts on a charge-$+q$ field, whereas $D_\mu^{(-q)}=\partial_\mu+iq A_\mu$ acts on its conjugate.
 With
\begin{equation}
 \widehat\phi(x):=\frac{\phi(x)}{|\phi(x)|}=e^{i\vartheta(x)},
\end{equation}
one obtains, wherever $\phi\neq0$,
\begin{equation}
\boxed{
W_\mu
 =\frac{i}{q}\widehat\phi^{\,*}D_\mu^{(+q)}[A]\widehat\phi
 =-\frac{i}{q}\widehat\phi\,D_\mu^{(-q)}[A]\widehat\phi^{\,*}.}
 \label{eq:KondoAbelianW}
\end{equation}
In the fixed-length theory ($\rho(x)=v$, or equivalently $|\phi(x)|=v/\sqrt2$), the gauge-invariant matter current, conserved on shell, is proportional to the same invariant vector \cite{Kondo2016PLB}:
$J^\mu \propto W_\mu$. 
In QRF language this identity is especially transparent: $\widehat\phi$ is precisely the $U(1)$ phase frame, and Eq.~\eqref{eq:KondoAbelianW} measures the gauge connection relative to that matter-built frame. Thus the operator-level gauge-independent massive field and the QRF relational connection are not merely analogous objects; in the Abelian complete-frame patch they are the same variable.

At the classical/tree level, suppose that $V(s)$, with $s=\phi^*\phi=\rho^2/2$, has a minimum at $\rho=v>0$, so $V'(v^2/2)=0$.
 The parameter $v$ here is a radial background scale, not the expectation value of a gauge-variant field.
	Then the \textbf{gauge-invariant radial expansion}, obtained by introducing the fluctuation field $H$ through
\begin{equation}
\boxed{
 \rho(x)=v+H(x)
}
\end{equation}
gives the kinetic term
\begin{align}
|D_\mu[A]\phi|^2
={}&\frac12(\partial_\mu H)^2
+\frac12q^2(v+H)^2W_\mu W^\mu \nonumber\\
={}&\frac12(\partial_\mu H)^2
+\frac12q^2v^2W_\mu W^\mu \nonumber\\
&+q^2vH W_\mu W^\mu
+\frac12q^2H^2W_\mu W^\mu.
\end{align}
which involves the mass term giving the mass  $m_W$ for $W$:
\begin{equation}
\boxed{
\frac12q^2v^2W_\mu W^\mu=\frac12m_W^2 W_\mu W^\mu, \quad
 m_W=qv.
}
\end{equation}
No gauge fixing and no assumption $\langle\phi\rangle\neq0$ have been used here. In particular, the unitary gauge condition $\vartheta=0$ has not been imposed at any stage of this derivation.

In the ordinary textbook, $\vartheta$ is called the Nambu--Goldstone field.
The conventional statement that ``the \textbf{Nambu--Goldstone mode} is absorbed by the gauge boson'' is simply a description obtained by decomposing the relational variable
$
 W_\mu=A_\mu-\frac1q\partial_\mu\vartheta
$
into $A_\mu$ and $\vartheta$.
Thus \textbf{the Higgs mass generation can be described completely using gauge-invariant variables alone.}
 Gauge-invariant descriptions of the Higgs mechanism of this type have long been known.
 The FMS work makes explicit that \textbf{the physical spectrum is to be constructed from gauge-invariant composite fields}.

The familiar Mexican-hat angular degeneracy is likewise not a physical degeneracy of gauge-inequivalent vacua. Pointwise in the Higgs fiber,
\begin{equation}
 \mathbb C/U(1)\simeq\mathbb R_{\ge0},
\end{equation}
hence the local gauge-invariant Higgs coordinate is the radial mode $\rho$. 
This is a local fiber statement and does not identify the full gauge-theory configuration space with $\mathbb R_{\ge0}$.

\subsection{Read $\vartheta$ not as a ``Nambu--Goldstone boson'' but as the orientation of an internal frame}
\label{subsec:abelian-frame}

Now we give the QRF interpretation of BEH mechanism.
Choose the phase of the Higgs field as QRF:
\begin{equation}
\boxed{
h(x)=e^{i\vartheta(x)}\in U(1).
}
\label{h-u1}
\end{equation}
\textbf{At each point $x$, this $h(x)$ specifies ``which direction in internal $U(1)$ space is to be taken as the reference direction.''}
Thus \textbf{$h(x)$ can be regarded as a local internal reference frame}.
The gauge transformation is given by
\begin{equation}
h(x)\mapsto e^{iq\alpha(x)}h(x),
\end{equation}
so the orientation of the Higgs frame seen from the outside is gauge dependent.
But, if the gauge field $A_\mu$ is rewritten as viewed from that Higgs frame $h$, the \textbf{relational Abelian gauge field} $A_\mu^h$ defined by
\begin{equation}
 \boxed{
 A_\mu^h(x):=A_\mu(x)+\frac{i}{q}h(x)^\dagger\partial_\mu h(x)
},
\end{equation}
has the explicit form for the choice (\ref{h-u1}) of $h$:
\begin{equation}
 \boxed{
A_\mu^h(x)=A_\mu(x)-\frac1q\partial_\mu\vartheta(x)=W_\mu(x)
},
\end{equation}
which is exactly gauge invariant.

Thus,
\begin{equation*}
\begin{aligned}
A_\mu={}&\text{connection relative to}\\
&\text{an external gauge frame} \ h.
\end{aligned}
\end{equation*}
whereas
\begin{equation*}
\begin{aligned}
W_\mu={}&\text{relational connection relative to}\\
&\text{the Higgs QRF }h.
\end{aligned}
\end{equation*}

The Higgs field itself becomes the relational field
\begin{equation}
\boxed{
 \phi^{h}(x)=h(x)^{-1}\phi(x)=e^{-i\vartheta(x)}\phi(x)=\frac{\rho(x)}{\sqrt2}.
}
\end{equation}
Thus the phase $\vartheta$ has not been discarded by gauge fixing; it has been absorbed into the definition of a gauge-invariant relational variable $\phi^h$.

The phase $\vartheta$ is not an independent observable angle but the orientation of the internal phase frame. $A_\mu$ by itself is also frame dependent. However,
$
 A_\mu-\frac1q\partial_\mu\vartheta
$
is the ``relative rotation rate of the connection with respect to the Higgs phase frame'' and is physical.

Thus \textbf{there is no need to regard the origin of the longitudinal polarization as the ``disappearance of a physical massless Nambu--Goldstone particle.''}
\textbf{
Rather, in (3+1)-dimensional spacetime, the three massive-vector polarizations appear as a result of reorganizing the gauge field $A_\mu$ and the QRF orientation $\vartheta$ into a single relational vector $W_\mu$.}

\subsection{Where did the Nambu--Goldstone boson go?}\label{subsec:goldstone}

The degree-of-freedom counting is unchanged by the relational description. In (3+1)-dimensions, a massless Abelian gauge field contributes two transverse polarizations, while the Higgs phase supplies one angular coordinate. Modulo the gauge redundancy, the pair $(A_\mu,\vartheta)$ is reorganized into the gauge-invariant massive vector
\begin{align}
 &\quad (A_\mu,\vartheta)/\text{gauge transformations}\nonumber\\
 \longleftrightarrow &\quad W_\mu=A_\mu-q^{-1}\partial_\mu\vartheta.
\end{align}
Thus the derivative of the frame orientation remains explicitly in the longitudinal sector of $W_\mu$; no physical massless Nambu--Goldstone particle has to exist first and then disappear. 
The gauge-invariant statement is simply that, in (3+1)-dimensions, the connection and the Higgs-frame orientation together provide the three polarizations of the massive relational vector.

\subsection{The meaning of unitary gauge also changes}
\label{subsec:unitary-abelian}

Ordinarily one sets the unitary gauge
\begin{equation}
\vartheta(x)\equiv 0,
\end{equation}
and writes
\begin{equation}
\phi(x)=\frac{v+H(x)}{\sqrt2}.
\end{equation}
Then the result
\begin{equation}
W_\mu(x)=A_\mu(x)
\end{equation}
is described as ``gauging away the Nambu--Goldstone boson.'' 

From the QRF viewpoint, the interpretation is instead that one adopts the Higgs frame itself as the internal coordinate reference, so that
the condition
$
\vartheta(x)=0
$
means choosing a frame in which
\begin{equation}
h(x)\equiv 1.
\end{equation}
In this frame,
$
 W_\mu(x)\equiv A_\mu(x).
$
The bare gauge field then appears massive because its coordinate representative coincides with the already gauge-invariant relational field.  Unitary gauge is therefore best viewed as the perspective in which the internal coordinate system is aligned with the Higgs QRF.

Thus unitary gauge is most economically described as the representation adapted to the Higgs QRF: setting $h=1$ makes the component representative of the already defined relational field $W_\mu$ coincide with the bare gauge potential $A_\mu$. This is a statement about representation, not the disappearance of a physical Nambu--Goldstone particle.

\subsection{The meaning of the mass term can also be understood relationally}

Even more interesting is the interpretation of the kinetic term
$
|D_\mu[A]\phi|^2.
$
$A_\mu$ is a connection that compares internal directions at different spacetime points, while the phase/orientation of the Higgs field defines an internal frame at each spacetime point. 
Therefore, \textbf{$D_\mu[A]\phi= \partial_\mu\phi-iqA_\mu\phi $ measures the discrepancy between the parallel transport of the frame predicted by the gauge connection
$A_\mu$ and the actual local orientation of the Higgs frame $h$}.

If these do not match, then $D_\mu[A]\phi\neq0$.  In the Abelian London limit $\rho(x)=v$, one has
\begin{equation}
 |D_\mu[A]\phi|^2
 =\frac12v^2(\partial_\mu\vartheta-qA_\mu)^2
 =\frac12q^2v^2W_\mu W^\mu.
\end{equation}
The coefficient $q^2v^2$ is positive and defines the mass-squared quadratic form in the internal/orbit direction.  The corresponding term in the Euclidean action, or in the Hamiltonian energy of physical modes, can be interpreted as a stiffness against relative misalignment between the Higgs frame orientation and the connection.  Increasing the radial scale $v$ increases this stiffness and the tree-level gap of the relational vector.

This is the Abelian prototype of the general BEH mechanism.

The mass term is the energy associated, in the Euclidean/Hamiltonian sense described above, with ``frame misalignment''.
The Higgs kinetic term
$
 |D_\mu[A]\phi|^2
$
has a very intuitive meaning from the QRF viewpoint.
This expresses gauge-invariant physics more directly than the explanation that ``a mass appears because the vacuum broke the gauge symmetry.''

\subsection{Relation to the Stueckelberg limit}

In the fixed-length or London limit,
$
 \rho(x)=v,
$
the Abelian theory becomes
\begin{align}
 \mathcal L
 ={}&-\frac14F_{\mu\nu}[A]F^{\mu\nu}[A]
 +\frac12v^2(\partial_\mu\vartheta-qA_\mu)^2
 \nonumber\\
 ={}&-\frac14F_{\mu\nu}[W]F^{\mu\nu}[W]
 +\frac12q^2v^2W_\mu W^\mu.
\end{align}
 The same gauge-invariant combination appears in Abelian Stueckelberg theory \cite{RueggRuiz-Altaba2004}.
  In the Abelian case, the Stueckelberg theory is itself perturbatively renormalizable and unitary, so the presence of a radial Higgs mode is not required merely to make the Abelian massive-vector theory renormalizable. 
From the QRF viewpoint, both have the same structure,
\begin{equation}
 \begin{aligned}
 &\quad\text{internal frame variable $h$}+\text{connection $A$}\\
  \longrightarrow&\quad\text{relational massive connection $W=A^h$}.
 \end{aligned}
\end{equation}

The sharper distinction arises in non-Abelian theories: a fixed-length Stueckelberg/nonlinear-sigma description is generally an effective theory, whereas a renormalizable Higgs theory contains additional radial dynamics that provides the standard ultraviolet completion. In the Higgs theory considered here, the frame variable is extracted from an already existing matter field, so QRF language clarifies the common relational structure without erasing this dynamical distinction.

\section{Complete BEH mechanism for non-Abelian case in QRF language}
\label{sec:nonabelian}

First, we consider the complete BEH mechanism for the non-Abelian gauge theory in QRF language.

\subsection{Complete Higgs frame}

The Lagrangian of the non-Abelian $G$ gauge--Higgs model is written in terms of the Higgs scalar field $\Phi$ and the gauge field $\mathscr A_\mu$ as
\begin{align}
 \mathscr L={}&-\frac12{\rm tr}\{\mathscr F_{\mu\nu}[\mathscr A]\mathscr F^{\mu\nu}[\mathscr A]\}\nonumber\\
 &+(D_\mu[\mathscr A]\Phi)^\dagger
 \cdot D^\mu[\mathscr A]\Phi 
 -V(\Phi^\dagger\cdot\Phi),\\
 \mathscr F_{\mu\nu}[\mathscr A]={}&
 \partial_\mu\mathscr A_\nu-\partial_\nu\mathscr A_\mu
 -ig[\mathscr A_\mu,\mathscr A_\nu],\\
  D_\mu[\mathscr A]={}&\partial_\mu-ig\mathscr A_\mu.
\end{align}
The Higgs field $\Phi(x)$ transforms under the local gauge transformation as
\begin{equation}
 \Phi(x)\mapsto \Omega(x) \Phi(x), \quad \Omega(x) \in G.
\end{equation}
Let the Higgs field $\Phi(x)$ determine a group-valued frame $h[\Phi](x)$ in a region where the gauge orientation can be resolved.  For a complete frame,  $h(x)$ transforms under the local gauge transformation as
\begin{equation}
 h[\Phi](x)\mapsto \Omega(x) h[\Phi](x), \quad \Omega(x) \in G.
\end{equation}
In the case of complete breaking, the relational connection $\mathscr{A}_\mu^h(x)$ is defined by  Eq.~\eqref{eq:relconnection}
\begin{equation}
 \mathscr{A}_\mu^h(x)
 =h(x)^\dagger\mathscr{A}_\mu(x) h(x)
 +\frac{i}{g}h(x)^\dagger\partial_\mu h(x),
\end{equation}
which is gauge invariant:
$
 \mathscr A_\mu^h(x)\mapsto\mathscr A_\mu^h(x) .
$

For the Higgs field as well, one defines a relational Higgs field from which the gauge orientation has been removed:
\begin{equation}
 \Phi^h(x) :=h(x)^{-1}\Phi(x).
\end{equation}
which is also gauge invariant: $\Phi^h(x)\mapsto \Phi^h(x)$.
The corresponding relational Higgs variable $\Phi^h(x)$ contains the radial and other gauge-invariant matter degrees of freedom but no redundant gauge orientation.

If the action is written in terms of $(\mathscr A_\mu^h,\Phi^h)$, the Nambu--Goldstone coordinates do not appear explicitly.
Thus the non-Abelian analogue of the Abelian reorganization is
\begin{equation}
 \boxed{
 (\mathscr{A}_\mu,\text{Higgs orientation}\ \hat\Phi)/G
 \quad\longrightarrow\quad
 \mathscr{A}_\mu^h
 }.
\end{equation}
The BEH mechanism is realized when the matter kinetic term produces a positive-semidefinite mass-squared quadratic form on the appropriate gauge-orbit components of the relational connection $\mathscr{A}_\mu^h$.

\subsection{$SU(2)$ gauge--fundamental Higgs model: a complete internal QRF}

Now we consider explicitly the complete BEH mechanism for the $SU(2)$ gauge--fundamental Higgs model in QRF language.

\subsubsection{Constructing a group-valued QRF from a Higgs doublet}

Let $\Phi(x)$ be an $SU(2)$ fundamental doublet field
\begin{equation}
 \Phi(x)=\begin{pmatrix}\phi_1(x)\\ \phi_2(x)\end{pmatrix}, \quad \phi_1(x), \phi_2(x) \in\mathbb{C}, 
\end{equation}
and define another $SU(2)$ doublet field $\widetilde\Phi(x)$ and an $SU(2)$ matrix field $\Theta(x)$: with the Pauli matrix$\sigma_2$, 
\begin{equation}
 \widetilde\Phi(x):=i\sigma_2\Phi(x)^*,
 \qquad
 \Theta(x):=(\widetilde\Phi(x),\Phi(x)),
\end{equation}
which obey
\begin{align}
 \Theta(x)^\dagger\Theta(x)
 & =\bigl(\Phi(x)^\dagger\cdot\Phi(x)\bigr)\mathbf 1_2,\\
 \det\Theta(x)&=\Phi(x)^\dagger\cdot\Phi(x).
\end{align}
Thus $\Theta$ itself is not, in general, an $SU(2)$ element; its normalized version is.
Defining the radial field by
\begin{equation}
 \rho(x)=\sqrt{2\Phi(x)^\dagger\cdot\Phi(x)} \ge 0,
\end{equation}
in the region or a patch with $\rho\neq0$,
one can construct the frame field
\begin{equation}
 \boxed{
 h(x)=\frac{\sqrt2}{\rho(x)}\Theta(x)= \frac{\Theta(x)}{\sqrt{\Phi(x)^\dagger \cdot\Phi(x)}} := \hat\Theta(x)
 \in SU(2),
 }
 \label{eq:higgsframe}
\end{equation}
which transforms under the local $SU(2)$ gauge transformation as 
\begin{equation}
 h(x)\mapsto \Omega(x) h(x).
\end{equation}
Thus $h$ is an internal QRF extracted from the Higgs field itself.
Hence \textbf{the Higgs doublet itself constructs a complete local color frame.}
This also coincides with a typical construction in the dressing field method \cite{FournelEtAl2012}.

Using the fixed vector $\xi_0$, 
the expression for $\Phi$ is extracted as
\begin{equation}
 \Phi(x)=\frac{\rho(x)}{\sqrt2}h(x)\xi_0, \quad \xi_0=\begin{pmatrix}0\\1\end{pmatrix}.
\end{equation}
The orientation $h$ and the radial magnitude $\rho$ have been separated without introducing a gauge-dependent vacuum direction.

\subsubsection{Relational gauge field and the exact mass term}

We use another notation $\mathscr{B}_\mu$ for the gauge-invariant relational gauge field $\mathscr{A}_\mu^h$ defined by Eq.~\eqref{eq:relconnection}
\begin{equation}
 \mathscr{A}_\mu^h
 =h^\dagger\mathscr{A}_\mu h
 +\frac{i}{g}h^\dagger\partial_\mu h
 =:\mathscr{B}_\mu=\mathscr{B}_\mu^a\frac{\sigma_a}{2}.
\end{equation}
The gauge-invariant relational matter field $\Phi^h$ is given by
\begin{equation}
 \Phi^h
 =h^{-1}\Phi= \frac{\rho}{\sqrt{2}} \xi_0 .
\end{equation}
The Lagrangian is rewritten in terms of the relational Higgs scalar field $\Phi^h$ and the relational gauge field $\mathscr B_\mu=\mathscr A_\mu^h$ as
\begin{align}
 \mathscr L={}&-\frac12{\rm tr}\{\mathscr F_{\mu\nu}[\mathscr B]
 \mathscr F^{\mu\nu}[\mathscr B]\} \nonumber\\
 &+(D_\mu[\mathscr B]\Phi^h)^\dagger
 \cdot\ D^\mu[\mathscr B]\Phi^h  
 -V(\Phi^h{}^\dagger\cdot\Phi^h),\\
 \mathscr F_{\mu\nu}[\mathscr B]={}&
 \partial_\mu\mathscr B_\nu-\partial_\nu\mathscr B_\mu
 -ig[\mathscr B_\mu,\mathscr B_\nu],\\
 D_\mu[\mathscr B]={}&\partial_\mu-ig\mathscr B_\mu.
\end{align}
The Lagrangian is gauge invariant under local $SU(2)$.
The Higgs kinetic term is exactly
\begin{align}
 (D_\mu[\mathscr{A}]\Phi)^\dagger \cdot D^\mu[\mathscr{A}]\Phi
 =& (D_\mu[\mathscr B]\Phi^h)^\dagger
 \cdot\ D^\mu[\mathscr B]\Phi^h \nonumber\\
 =&\frac12(\partial_\mu\rho)^2
 +\frac{g^2}{8}
 \rho^2\mathscr{B}_\mu^a\mathscr{B}^{a\mu}
 .
 \label{eq:su2kinetic}
\end{align}

Then it is rewritten in terms of the gauge-invariant radial Higgs scalar field $\rho$ and the relational gauge field $\mathscr B_\mu=\mathscr A_\mu^h$ as
\begin{align}
 \mathscr L={}&-\frac12 {\rm tr}\{\mathscr F_{\mu\nu}[\mathscr B]
 \mathscr F^{\mu\nu}[\mathscr B]\}
 +\frac12(\partial_\mu\rho)^2
 \nonumber\\
 &+\frac{g^2}{8}\rho^2\mathscr B_\mu^a\mathscr B^{a\mu}
 -V\left(\frac12\rho^2\right).
\end{align}
Therefore, near  a gauge-invariant radial minimum $\rho(x)=v$, $\mathscr B_\mu$ has the mass $m_{\mathscr B}$ in the tree level:
\begin{equation}
\frac{g^2}{8}v^2\mathscr B_\mu^a\mathscr B^{a\mu}
=\frac{1}{2}m_{\mathscr B}^2\mathscr B_\mu^a\mathscr B^{a\mu}, \quad m_{\mathscr B}=\frac{gv}{2}.
\end{equation}

The fixed-length version $\rho(x)=v$ makes the point especially sharply: even when the radial Higgs particle $H$ is frozen out, the angular matter field still provides a complete frame and the vector mass term remains.
Here again, only a gauge-invariant nonzero scale in the radial field $\rho$ is required, not an orientation of $\langle\Phi\rangle$. In a fixed-length model, vector mass generation occurs even after the Higgs particle $H$ itself is removed.

\subsection{
Kondo's fundamental massive mode as a complete-QRF relative field
}

The above result is consistent with the gauge-independent BEH construction based on a fixed-length fundamental scalar:
In the gauge-independent formulation of BEH mechanism in the $SU(2)$ gauge-fundamental scalar model,
using the normalized Higgs matrix $\hat \Theta(x)$
a \textbf{gauge-covariant massive vector mode} $\mathscr{W}_\mu(x)$ was introduced by \cite{Kondo2018}
\begin{equation}
\boxed{
 \mathscr{W}_\mu(x)
 :=\frac{i}{g}(D_\mu[\mathscr{A}]\hat \Theta(x))\hat \Theta(x)^\dagger .
}
\end{equation}
The field $\hat \Theta(x)$ transforms under the local gauge transformation as
\begin{equation}
 \hat \Theta(x)(x)\mapsto \Omega(x)\hat \Theta(x)(x),
 \qquad \Omega(x)\in SU(2).
\end{equation}
Then the field $\mathscr{W}_\mu(x)$ transforms adjoint-covariantly under the local gauge transformation as :
\begin{equation}
 \mathscr{W}_\mu(x)\mapsto \Omega(x)\mathscr{W}_\mu(x)\Omega(x)^\dagger,
 \qquad \Omega(x)\in SU(2).
\end{equation}
In addition, a \textbf{gauge-invariant massive vector mode} $\tilde{\mathscr{W}}_\mu(x)$ was also introduced by \cite{Kondo2018}
\begin{equation}
\boxed{
 \tilde{\mathscr{W}}_\mu(x)
 :=\hat \Theta(x)^\dagger\mathscr{W}_\mu(x) \hat \Theta(x)
 =\frac{i}{g}\hat \Theta(x)^\dagger D_\mu[\mathscr{A}] \hat \Theta(x),
}
\end{equation}
which is invariant under the gauge transformation: 
\begin{equation}
 \tilde{\mathscr{W}}_\mu(x)\mapsto \tilde{\mathscr{W}}_\mu(x),
 \qquad \Omega(x)\in SU(2).
\end{equation}

We can understand these fields using the Higgs frame in the QRF description.
In fact, defining the Higgs-frame field 
\begin{align}
h(x) := \hat \Theta(x) \in SU(2),
\end{align}
we find that
the gauge-invariant massive mode $\tilde{\mathscr W}_\mu:=h^\dagger\mathscr W_\mu h$  of Ref.~\cite{Kondo2018} is the relational field of the gauge-covariant massive mode $\mathscr{W}_\mu$ and is equal to the relational field $\mathscr{A}_\mu^h(x)=h^\dagger\mathscr{A}_\mu h+\frac{i}{g}h^\dagger\partial_\mu h$ of the original gauge field $\mathscr{A}_\mu(x)$:
\begin{align}
 &\boxed{
 \tilde{\mathscr{W}}_\mu(x)=h(x)^\dagger\mathscr{W}_\mu(x) h(x)=\mathscr{A}_\mu^h(x)
=\frac{i}{g}h(x)^\dagger D_\mu[\mathscr{A}] h(x)
 } .
\end{align}
The gauge-covariant massive mode $\mathscr{W}_\mu$ and the gauge-invariant relational field $\mathscr{A}_\mu^h$ contain the same physical information, since they have the same mass term:
\begin{equation}
 \operatorname{tr}(\mathscr{W}_\mu\mathscr{W}^\mu)
 =\operatorname{tr}(\tilde{\mathscr{W}}_\mu\tilde{\mathscr{W}}^\mu)=\operatorname{tr}(\mathscr{A}_\mu^h\mathscr{A}^h{}^\mu).
\end{equation}

Classically, on a regular complete-frame patch, the Higgs-frame components of $\mathscr A_\mu^h=\mathscr B_\mu$ are invariant under the original local $SU(2)$ redundancy.  In the quantum theory this statement is to be understood through renormalized gauge-invariant composite interpolators, such as $\mathcal O_\mu^a\propto\rho^2\mathscr B_\mu^a$ (see next section VI), rather than by assuming that the normalized nonpolynomial field $\mathscr B_\mu$ is automatically a renormalized pointlike operator.

The separation of the variable $\mathscr{W}_\mu(x)$ has the specific meaning in understanding quark confinement.
The original gauge field is decomposed as \cite{Kondo2018}
\begin{align}
 &\mathscr{A}_\mu=\mathscr{V}_\mu+\mathscr{W}_\mu, \nonumber\\
 & \mathscr{V}_\mu:=\frac{i}{g}h\partial_\mu h^\dagger, \quad
 \mathscr{W}_\mu(x)
 :=\frac{i}{g}(D_\mu[\mathscr{A}]h(x))h(x)^\dagger
\end{align}
which obey the gauge transformation:
\begin{align}
  \mathscr{V}_\mu(x) \mapsto&
 \Omega(x)\mathscr{V}_\mu(x)\Omega(x)^{-1}
 +\frac{i}{g}\Omega(x)\partial_\mu\Omega(x)^{-1},
 \nonumber\\
  \mathscr{W}_\mu(x) \mapsto&
 \Omega(x)\mathscr{W}_\mu(x)\Omega(x)^{-1}, \qquad \Omega(x)\in G.
\end{align}
Here $\mathscr{V}_\mu$ is the connection associated with the QRF orientation of the Higgs frame, while $\mathscr{W}_\mu$ is the relative massive mode.
 Although $\mathscr{V}_\mu$ looks perturbatively like a pure gauge, it can retain topology/defects in sectors in which $h$ cannot be chosen globally single-valued and smooth. This point is directly connected with the global obstruction of the QRF.

The above considerations suggest \cite{Kondo2018} that all comonents of the field
$\mathscr{W}_\mu(x)$ become massive by the BEH mechanism and are expected to be decoupled in the infrared, then $\mathscr{V}_\mu(x)$ play the dominant role in quark confinement by incorporating the effects of topological defects.

\section{
Relation to the FMS mechanism: exact relational identity and perturbative pole matching
}
\label{sec:fms}

The core of Fr\"ohlich--Morchio--Strocchi (FMS) is to read the physical spectrum from gauge-invariant composite operators \cite{FMS1980,FMS1981,Maas2019}.

The QRF variables make their relation to ordinary massive-vector fields especially transparent.

For the $SU(2)$ fundamental Higgs $\Phi(x)$, define
\begin{equation}
 \mathcal O_H(x):=\Phi(x)^\dagger\cdot\Phi(x)=\frac{1}{2}\rho(x)^2.
\end{equation}
This is the gauge-invariant scalar operator.

As a FMS vector operator, take a vector composite carrying a custodial index $a$ as
\begin{equation}
\boxed{
 \mathcal O_\mu^a(x)
 :=i\,\mathrm{tr}\!\left[
 \sigma^a\Theta(x)^\dagger D_\mu[\mathscr A]\Theta(x)
 \right].
}
\end{equation}
Here the index $a$ is not a surviving local gauge index.  The local $SU(2)$ gauge index has been relationalized by the Higgs frame; $a$ labels the physical custodial/global orientation of this gauge-invariant vector composite (equivalently, the components referred to the Higgs frame).  This is why $\mathcal O_\mu^a$ can be locally gauge invariant while still forming a nontrivial multiplet under the custodial symmetry.
A direct calculation gives
\begin{equation}
 \boxed{
 \mathcal O_\mu^a(x)
 =\frac{g}{2}\rho(x)^2\mathscr{B}_\mu^a(x)
 }
 \label{eq:FMSQRF}
\end{equation}
Thus the FMS vector operator $\mathcal O_\mu^a(x)$ is precisely the relational massive vector $\mathscr{B}_\mu$ multiplied by a gauge-invariant radial weight $\rho$.
Equation~\eqref{eq:FMSQRF} is, with the definitions used here, an exact classical identity between a gauge-invariant composite operator and the relational field multiplied by the gauge-invariant radial weight.

Expanding only the gauge-invariant radial variable,
\begin{equation}
 \rho(x)=v+H(x),
\end{equation}
one finds
\begin{align}
 \mathcal O_H(x)
 &=\frac{1}{2}v^2+vH(x)+\frac12H(x)^2,
 \label{eq:fms-expand-scalar}\\
 \mathcal O_\mu^a(x)
 &=\frac{g}{2}v^2\mathscr{B}_\mu^a(x)
+gvH(x)\mathscr{B}_\mu^a(x)
 +\frac{g}{2}H(x)^2\mathscr{B}_\mu^a(x).
 \label{eq:fms-expand-vector}
\end{align}
The FMS pole statement is a separate perturbative step: after expanding the composite operator in a Standard-Model-like Higgs regime, the leading term of its two-point function has the same pole as the corresponding gauge-fixed elementary-vector propagator, up to the usual assumptions and higher composite contributions. Thus the physical vector excitation is represented by a gauge-invariant composite operator; the gauge-fixed elementary $W$ field is an efficient interpolating field whose pole agrees with that composite channel at leading FMS order in the regime where the expansion is controlled. For more general gauge groups and Higgs representations, this simple spectral matching need not persist \cite{Maas2019}.

The spectral characterization depends on the physical channel.  A stable neutral one-particle state may be identified by an isolated pole, or equivalently by the corresponding spectral gap.  An unstable state is characterized by resonance data, conventionally a complex pole after the appropriate analytic continuation.  A residual-charged sector coupled to a massless gauge field requires an infrared-sensitive dressed spectral analysis: because of the soft gauge-field cloud, the relevant spectral measure need not contain an isolated one-particle pole and may instead exhibit an infraparticle-type threshold or branch-point singularity \cite{Buchholz1986}.  Accordingly, the classical/tree-level BEH mass parameter, a stable pole mass, a resonance pole, and a nonperturbative mass gap are kept conceptually distinct below.

The same logic is visible in the Abelian current as the conserved Noether current associated with $U(1)$ symmetry:
\begin{equation}
 J_\mu
 =i\phi^*D_\mu\phi-i(D_\mu\phi)^*\phi,
\end{equation}
which is gauge invariant.
Substituting the polar representation (\ref{polar}) of $\phi$ gives, up to normalization,
\begin{equation}
 J_\mu\propto\rho^2(qA_\mu-\partial_\mu\vartheta)
 \propto {\rho^2W_\mu}.
\end{equation}
The physical vector pole is therefore naturally associated with a gauge-invariant relational composite $W_\mu$ rather than the bare gauge potential $A_\mu$ by itself.

In quantum theory, physical observables must be invariant under the local gauge redundancies generated by the Gauss constraint. Physical states may nevertheless carry genuine asymptotic or global charges, in which case a local charged field by itself is not a physical operator and an appropriate Dirac, Wilson-line, or asymptotic dressing is required. The gauge-invariant formulation of FMS addresses the neutral composite spectrum directly.

In the FMS description, the physical massive-vector excitation is represented by a gauge-invariant composite.
The physical vector excitation is interpolated by a gauge-invariant composite such as $\mathcal O_\mu^a$ in Eq.~\eqref{eq:FMSQRF}, rather than by the bare gauge potential itself.  
Expanding $\rho=v+H$ in the BEH regime, the leading contribution to the two-point function of the gauge-invariant composite reproduces the familiar massive-vector pole under the usual FMS assumptions.

At the quantum level the renormalized custodial vector composite and its Ward identities provide the appropriate operator framework.  In particular, the Ward-identity analysis separates the physical propagating spin-one channel from longitudinal or constraint contributions that do not represent an additional propagating particle pole \cite{DudalEtAl2022}.  This distinction is useful when relating the frame-orientation sector of the relational description to the physical vector spectrum.
\textbf{
In QRF/dressing language, the gauge-fixed elementary massive vector is therefore interpreted as the leading perturbative representative of a gauge-invariant composite excitation described relative to the Higgs frame.
}
In this sense, the FMS approach and the QRF/dressing approach express very closely related physics. The former emphasizes gauge-invariant composite operators and their spectrum, whereas the latter emphasizes the structure of the internal frame and relational variables.

\section{Partial BEH mechanism for non-Abelian case in QRF language}
\label{sec:partial}

Next,  we consider the partial BEH mechanism for the non-Abelian gauge theory in QRF language.


\subsection{A coset frame rather than a complete frame}

For a general Higgs representation, let $\Phi_0$ be a reference configuration and define its \textbf{stabilizer} or \textbf{stability subgroup}

\begin{equation}
\boxed{
 H=\{g\in G\mid R(g)\Phi_0=\Phi_0\} \subset G.
}
\end{equation}
The Higgs orientation determines a point of the orbit $G/H$, rather than a unique element of $G$.  It is therefore most precisely regarded as a \textbf{coset-valued reference orientation}.  A local lift $h\in G$ can be chosen, but only up to the right action of the stabilizer $H$; consequently the Higgs data do not determine a complete basis in the coset tangent space.

If a local section $h[\Phi](x)$ is chosen, its choice has the residual ambiguity
\begin{equation}
 h(x)\sim h(x)k(x),
 \quad k(x)\in H, \quad h(x) \in G .
\end{equation}
Accordingly, the relational connection $\mathscr A_\mu^h$ is not a completely gauge-invariant field and
 transforms as a connection under the residual $H$ gauge transformation.
If the  frame $h(x)$ transforms under a gauge transformation and a compensating change of representative as
\begin{equation}
h(x)\mapsto \Omega(x)h(x)r(x)^{-1},
\qquad
r(x)\in H,
\end{equation}
the relational connection transforms under the residual group as
\begin{equation}
 \mathscr{A}_\mu^h(x)
 \mapsto
 r(x)\mathscr{A}_\mu^h(x) r(x)^{-1}
 +\frac{i}{g}r(x)\partial_\mu r(x)^{-1}.
\end{equation}
Thus the original $G$ redundancy has been relationalized only to the extent supplied by the coset-valued Higgs data: a residual $H$ gauge freedom remains.  Geometrically, the Higgs configuration selects a point of the orbit $G/H$ and the associated orbit-tangent/coset subspace, while a basis within that subspace is defined only up to the residual $H$ action.  The Higgs kinetic term generates the BEH mass-squared form along these coset directions, whereas the $H$ directions remain a residual gauge connection.  The conventional ``broken generators'' therefore denote orbit directions on which Higgs matter supplies reference orientation data, not a literal physical breaking of the original local $G$ redundancy.

\subsection{The mass matrix is the metric on the gauge orbit}

\paragraph{Proposition (Higgs orbit metric and BEH mass matrix).}
Let the scalar kinetic term carry a positive internal metric and let $v\Phi_0$ be a reference configuration with stabilizer $H$.  The tree-level vector mass-squared quadratic form is the metric induced on the tangent space to the gauge orbit through $\Phi_0$ by the orbit map $G\to G\!\cdot\!\Phi_0$.  Its kernel is the stabilizer algebra $\mathfrak h$; if the stabilizer is precisely $H$, then its rank is $\dim(G/H)$.  Equations~\eqref{eq:massmatrix} and the rank relation (\ref{M2}) are the component form of this geometric statement.

For the compact groups of primary interest here, choose an $H$-invariant complement $\mathfrak m$ of the stabilizer algebra $\mathfrak h$ of the total algebra $\mathfrak g$ so that
\begin{equation}
 \mathfrak g=\mathfrak h\oplus\mathfrak m.
\end{equation}
Then $\mathfrak m$ represents the tangent directions to the \textbf{orbit} $G/H$ at the reference configuration and contains the directions for which the Higgs frame can generate a classical BEH mass term, whereas $\mathfrak h$ contains the residual gauge directions.

At the classical/tree level, the Higgs kinetic term induces a positive semidefinite metric on the gauge orbit in field space. For a radial background $v\Phi_0$ with positive scalar kinetic metric, and with $H$ the stabilizer of the reference configuration $\Phi_0$, the vector mass matrix can be written as
\begin{equation}
 (M^2)_{ab}
 =2g^2v^2\,\operatorname{Re}
 \left[(T_a\Phi_0)^\dagger(T_b\Phi_0)\right],
 \label{eq:massmatrix}
\end{equation}
where the normalization of $v$ is adjusted to the convention for $\Phi_0$. A generator satisfying $T_a\Phi_0=0$ belongs to the stabilizer algebra $\mathfrak h$ and is a zero direction of this Higgs-generated mass matrix. Under these assumptions, and for a reference configuration whose stabilizer is precisely $H$,
\begin{equation}
 \operatorname{rank}M^2=\dim(G/H).
 \label{M2}
\end{equation}
\textbf{In QRF language, the Higgs matter supplies a coset-valued reference orientation on $G/H$, with a local lift defined only up to $H$.  The corresponding coset components of $\mathscr A_\mu^h$ acquire the Higgs-generated mass-squared form, whereas the residual $H$ connection receives no mass term from the Higgs kinetic term alone.}
\textbf{This statement concerns the classical Higgs-generated mass matrix and does not exclude other quantum or nonperturbative mass-gap effects in the residual sector.}

\subsection{$SU(2)$ gauge--adjoint Higgs model: partial internal QRF and residual $U(1)$}

Now we consider explicitly the partial BEH mechanism for the $SU(2)$ gauge--adjoint Higgs model in QRF language.  We use a single matrix/vector notation throughout this section.  Let
\begin{equation}
 \mathscr X(x)=\mathscr X^a(x)T_a,\quad
 \bm X\!\cdot\!\bm Y:=2\operatorname{tr}(\bm X\bm Y),
 \quad T_a=\frac{\sigma_a}{2}.
\end{equation}
The Lagrangian is written as
\begin{align}
 \mathscr L={}&-\frac12\operatorname{tr}\!\left\{\mathscr F_{\mu\nu}[\mathscr A]\mathscr F^{\mu\nu}[\mathscr A]\right\}
 \nonumber\\
 &+\frac12\mathscr D_\mu[\mathscr A]\bm\phi\!\cdot\!\mathscr D^\mu[\mathscr A]\bm\phi
 -V(\bm\phi\!\cdot\!\bm\phi),\\
 \mathscr F_{\mu\nu}[\mathscr A]:={}&\partial_\mu\mathscr A_\nu-\partial_\nu\mathscr A_\mu
 -ig[\mathscr A_\mu,\mathscr A_\nu],\\
 \mathscr D_\mu[\mathscr A]\bm\phi:={}&\partial_\mu\bm\phi-ig[\mathscr A_\mu,\bm\phi].
\end{align}
Writing 
\begin{equation}
\bm\phi(x)=\rho(x)\bm n(x)
\end{equation} 
away from $\rho=0$, the normalized color direction satisfies
\begin{equation}
 \bm n(x)=n^a(x)\frac{\sigma_a}{2},
 \qquad \bm n(x)\cdot\bm n(x)=1.
\end{equation}

Locally one can write
\begin{equation}
 \bm n(x)=h(x)\frac{\sigma_3}{2}h(x)^\dagger,
\end{equation}
but $h$ has the residual $U(1)$ ambiguity
\begin{equation}
 h(x)\sim h(x) e^{i\beta(x)\sigma_3/2}.
\end{equation}
For an $SU(2)$ adjoint Higgs, the normalized color direction $\bm n$ determines only a point in
\begin{equation}
 SU(2)/U(1)\simeq S^2.
\end{equation}

Thus $\bm n$ selects the two-dimensional tangent/coset subspace orthogonal to itself, but it does not select a unique basis within that plane: rotations of that basis are precisely the residual $U(1)$ ambiguity. The two coset components receive the classical Higgs-generated mass term, whereas the residual $U(1)$ connection receives no such term. 
This is the \textbf{partial-QRF interpretation of adjoint Higgsing without assigning physical breaking to the local $SU(2)$ redundancy.}

At the quantum level, however, the existence of a Higgs-induced mass term for the residual-$U(1)$-charged coset field is logically distinct from the existence of a local gauge-invariant operator that interpolates that charged excitation. Recent analysis of local gauge-invariant vector operators in the adjoint $SU(2)$ Higgs model makes this distinction explicit by using equations of motion, BRST structure, and Ward identities to separate propagating from nonpropagating local vector channels \cite{PeruzzoAdjoint2026}. A residual-$U(1)$-charged massive excitation therefore requires an additional physical dressing and is generally represented by a nonlocal or asymptotic operator.
This is a general partial-frame principle used throughout the paper:
\begin{equation}
 \begin{aligned}
 &\text{Higgs-induced stiffness/mass matrix}\\
 &\hspace{1.5em}\not\Rightarrow
 \text{ local physical charged interpolator}.
 \end{aligned}
\end{equation}
The adjoint-Higgs coset modes and the electroweak $W^\pm$ are two realizations of the same distinction.

\subsection{Kondo's adjoint massive mode as a partial-QRF relative field}

The gauge-\textit{independent} $SU(2)$ adjoint-Higgs construction of Ref.~\cite{Kondo2016PLB,Kondo2017EPJ}, makes the partial-frame structure explicit at the operator level. For a radially fixed adjoint field $\bm\phi(x)\cdot\bm\phi(x)=v^2$ \begin{equation}
\bm n(x):=\frac{\bm\phi(x)}{v}, \quad
\bm n(x)\cdot\bm n(x)=1,
\end{equation} 
the gauge-covariant massive mode is introduced \cite{Kondo2016PLB}
\begin{equation}
\boxed{
 \mathscr W_\mu(x)
 :=-\frac{i}{g}\,[\bm n(x),\mathscr D_\mu[\mathscr A]\bm n(x)].
}
 \label{eq:adjointKondoW}
\end{equation}
For the adjoint scalar field with the gauge transformation
\begin{align}
\bm\phi(x) \mapsto 
 \Omega(x)\bm\phi(x)\Omega(x)^{-1} \qquad \Omega(x)\in G,
\end{align}
and hence
\begin{align}
\bm n(x) \mapsto 
 \Omega(x)\bm n(x)\Omega(x)^{-1} \qquad \Omega(x)\in G,
\end{align}
Therefore, $\mathscr W_\mu(x)$ obeys an adjoint gauge transformation, 
\begin{align}
\mathscr{W}_\mu(x) \mapsto 
 \Omega(x)\mathscr{W}_\mu(x)\Omega(x)^{-1} \qquad \Omega(x)\in G.
\end{align}
Note that $\mathscr W_\mu(x)$ is orthogonal to the color direction,
\begin{equation}
 \mathscr W_\mu(x)\cdot\bm n(x)=0,
\end{equation}
which means that $\mathscr W_\mu(x)$ has only the coset components. This  is a gauge-invariant condition.  

In general, the connection is decomposed into the diagonal (stabilizer) and off-diagonal (coset) parts:
\begin{align}
 &\mathscr B_\mu
 = \mathscr B_\mu^\parallel+\mathscr B_\mu^\perp , \nonumber\\
 &\mathscr B_\mu^\parallel=B_\mu^3T_3, \nonumber\\
 &\mathscr B_\mu^\perp=B_\mu^1T_1+B_\mu^2T_2 = [T_3, [T_3, \mathscr B_\mu]] . 
\end{align}
Then $\mathscr B_\mu^\parallel$ is the residual $U(1)$ connection, whereas $\mathscr B_\mu^\perp$ is a coset $SU(2)/U(1)$ field transforming homogeneously under the residual $U(1)$. 

The massive mode $\mathscr W_\mu(x)$ is gauge-covariant. 
By introducing the frame $h$ through
\begin{equation}
\bm n(x)=h(x)T_3h(x)^\dagger ,
\end{equation}
 the gauge-\textit{invariant} relational massive mode $\tilde{\mathscr{W}}_\mu$  is obtained (see Appendix A for the derivation)
\begin{align}
 \tilde{\mathscr W}_\mu(x) :=& h(x)^\dagger\mathscr W_\mu(x) h(x) \nonumber\\
 =&
 (\mathscr{A}_\mu^h(x))^\perp := [T_3,[T_3,\mathscr{A}_\mu^h(x)]].
 \label{eq:W=Bperp}
\end{align}
Hence the adjoint Higgs field realizes the partial relationalization:
\begin{equation}
 SU(2)
 \quad\longrightarrow\quad
 \underbrace{SU(2)/U(1)}_{\text{frame supplied}}
 +
 \underbrace{U(1)}_{\text{unfixed residual frame}}
 \label{eq:partialrelationalization}
\end{equation}

The Higgs kinetic term for the exact fixed-length scalar becomes the mass term for the coset $SU(2)/U(1)$ field:
\begin{align}
 &\frac12 \mathscr D_\mu[\mathscr A]\bm\phi(x)\cdot \mathscr D^\mu[\mathscr A]\bm\phi(x)
 = \frac12(gv)^2\,\mathscr W_\mu(x)\cdot\mathscr W^\mu(x)
 \nonumber\\
 =&\frac12(gv)^2\,(\mathscr{A}_\mu^h(x))^\perp\cdot(\mathscr{A}_\mu^h(x))^\perp.
 \label{eq:adjointKondoMass}
\end{align}
Therefore, at the radial minimum $\rho=v$, the masses of coset $SU(2)/U(1)$ and $U(1)$ parts are respectively given by
\begin{equation}
 m_{SU(2)/U(1)}=gv,
 \qquad
 m_{U(1)}=0.
\end{equation}

The original connection can then be decomposed as
\begin{align}
 \mathscr A_\mu&=\mathscr V_\mu+\mathscr W_\mu,\nonumber\\
 \mathscr V_\mu&=c_\mu\bm n+\frac{i}{g}[\bm n,\partial_\mu\bm n], \quad
 c_\mu :=\mathscr A_\mu\cdot\bm n,\nonumber\\
 \mathscr W_\mu
 &= -\frac{i}{g}\,[\bm n,\mathscr D_\mu[\mathscr A]\bm n].
\end{align}

The QRF interpretation sharpens the status of these variables. The color direction $\bm n$ is a partial internal QRF taking values in $SU(2)/U(1)$. The field $\mathscr W_\mu$ is the coset-relative motion of the connection with respect to that partial frame and acquires the Higgs stiffness in Eq.~\eqref{eq:adjointKondoMass}. By contrast, $\mathscr V_\mu$ is the connection compatible with the partial frame and retains the residual $U(1)$ gauge freedom. Accordingly, $\mathscr W_\mu$ itself is gauge covariant rather than a set of gauge-invariant local components; after choosing a local lift $h$ of $\bm n=h(\sigma_3/2)h^\dagger$, its Higgs-frame components are only residual-$U(1)$ covariant. This is precisely what one expects from an incomplete QRF.

\section{Electroweak BEH mechanism of the Standard Model in QRF language}
\label{sec:electroweak}

Finally, we consider the electroweak BEH mechanism of the Standard Model in QRF language.

The Lagrangian of the $SU(2)_L\times U(1)_Y$ gauge--Higgs model is written in terms of the fundamental Higgs scalar field $\Phi$, the $SU(2)_L$ connection $\mathscr A_\mu$, and the hypercharge connection $B_\mu^Y$ as
\begin{align}
 \mathscr L={}&-\frac12 {\rm tr}\{\mathscr F_{\mu\nu}[\mathscr A]\mathscr F^{\mu\nu}[\mathscr A]\}
 -\frac14 B_{\mu\nu}^{Y}B_Y^{\mu\nu}
 \nonumber\\
 &+(D_\mu[\mathscr A,B^Y]\Phi)^\dagger D^\mu[\mathscr A,B^Y]\Phi -V(\Phi^\dagger\Phi),\\
 \mathscr F_{\mu\nu}[\mathscr A]:={}&\partial_\mu\mathscr A_\nu-\partial_\nu\mathscr A_\mu
 -ig[\mathscr A_\mu,\mathscr A_\nu],\\
 B_{\mu\nu}^{Y}:={}&\partial_\mu B_\nu^Y-\partial_\nu B_\mu^Y,\\
 D_\mu[\mathscr A,B^Y]\Phi:={}&\left(\partial_\mu-ig\mathscr A_\mu
 -i\frac{g'}{2}B_\mu^Y\right)\Phi.
\end{align}
Let the Higgs doublet be
\begin{equation}\label{eq:display-28}
\Phi=\begin{pmatrix}\phi^+\\\phi^0\end{pmatrix}.
\end{equation}
The Higgs-field potential is
\begin{equation}\label{eq:display-11}
V(\Phi^\dagger\Phi)=-\mu^2\Phi^\dagger\Phi
+\lambda(\Phi^\dagger\Phi)^2,
\qquad \mu^2,\lambda>0.
\end{equation}
Its minimum condition is
\begin{equation}\label{eq:display-12}
\Phi^\dagger\Phi=\frac{v^2}{2},
\qquad v^2:=\frac{\mu^2}{\lambda}.
\end{equation}

Let the two independent local gauge transformations be
\begin{equation}\label{eq:display-4}
 \begin{aligned}
 \Omega(x)&=\exp\!\bigl(i\alpha^a(x)T^a\bigr)\in SU(2)_L,\\
 g_Y(x)&:=e^{i\beta(x)Y}\in U(1)_Y.
 \end{aligned}
\end{equation}
For the Standard Model Higgs doublet $Y_\Phi=1/2$, and hence
\begin{equation}\label{eq:old-2-1}
 \Phi(x)\mapsto \Omega(x)g_Y(x)\Phi(x)
 =\Omega(x)e^{i\beta(x)/2}\Phi(x).
\end{equation}

\subsection{$SU(2)_L$ Higgs frame and residual $U(1)$}
\label{subsec:ew-frame}

Let the Standard Model Higgs doublet $\Phi$ have hypercharge $Y=1/2$, and write

\begin{align}
&\boxed{
 \Theta(x):=(\widetilde\Phi(x),\Phi(x))
 =\frac{\rho(x)}{\sqrt2}h(x),
} 
 \nonumber\\
 \Leftrightarrow 
&\boxed{
 h(x)=\frac{\sqrt2}{\rho(x)}\Theta(x), \quad
 \rho(x)=\sqrt{2\Phi(x)^\dagger\cdot\Phi(x)}.
}
\end{align}
Here the Higgs doublets $\Phi(x),\widetilde\Phi(x)$ are written in terms of the frame field $h(x)$:
\begin{align}
 \Phi(x)&=\frac{1}{\sqrt2}\rho(x)h(x)\xi_0, \quad \xi_0 =\begin{pmatrix}0\\1\end{pmatrix} , \nonumber\\
 \widetilde\Phi(x)&=\frac{1}{\sqrt2}\rho(x)h(x) \widetilde\xi_0, \quad  \widetilde\xi_0 =\begin{pmatrix}1\\0\end{pmatrix} .
\end{align}

Under $SU(2)_L\times U(1)_Y$ transformations, the Higgs-frame matrix transforms as
\begin{align}
 h(x)&\mapsto\Omega(x)h(x)r_\beta(x)^{-1},\\
 r_\beta(x)&:=e^{i\beta(x)\sigma_3/2}\in U(1)_R\subset SU(2)_R.
\end{align}
Here $r_\beta$ 
is the right action on the Higgs-frame matrix induced by the original $U(1)_Y$ transformation.  The $SU(2)_L$ action from the left is relationalized by the Higgs QRF, while this induced right action remains as the residual local structure.  
On relational fields it is generated by the electromagnetic charge $Q=T^3+Y$, and is therefore identified locally with $U(1)_{\rm em}$.

This identification is a statement about the local Lie-algebra action.  The global form of the electroweak gauge group can involve a discrete quotient and depends on the complete matter content and hypercharge normalization; no stronger global-group statement is assumed here.

For the $SU(2)_L$ connection $\mathscr{A}_\mu$ and the hypercharge connection $B_\mu^Y$,  define the relational $SU(2)_L$ connection $\mathscr{A}_\mu^h(x)$ by
\begin{equation}
 \mathscr{A}_\mu^h(x)
 =h(x)^{-1}\mathscr{A}_\mu(x) h(x)
 +\frac{i}{g}h(x)^{-1}\partial_\mu h(x):=\mathscr{B}_\mu(x).
\end{equation}

Writing the residual transformation as $h\to\Omega_L h r^{-1}$ with $r_\beta=e^{i\beta\sigma_3/2}$ gives
\begin{align}
 \mathscr B_\mu&\mapsto r_\beta\mathscr B_\mu r_\beta^{-1}+\frac{i}{g}r_\beta\partial_\mu r_\beta^{-1}, \quad
 \mathscr B_\mu^3&\mapsto \mathscr B_\mu^3+\frac{1}{g}\partial_\mu\beta,\nonumber\\
 B_\mu^Y&\mapsto B_\mu^Y+\frac{1}{g'}\partial_\mu\beta.
\end{align}

By using the relation following from the definition of the relational connection $\mathscr{B}_\mu$,
\begin{align}
 h^{-1}\left(\partial_\mu-ig\mathscr{A}_\mu\right)h
 =-ig \mathscr{B}_\mu.
\end{align}
the covariant derivative becomes
\begin{align}
 D_\mu[\mathscr A,B^Y]\Phi(x)
={}&\frac{1}{\sqrt2}h(x)
 \Bigl[(\partial_\mu\rho(x))\xi_0
 \nonumber\\
 &-i\rho(x)
 \left(g\mathscr B_\mu(x)+\frac{g'}{2}B_\mu^Y(x)\mathbf 1_2\right)
 \xi_0\Bigr].
\end{align}
Using explicitly $\xi_0 =\begin{pmatrix}0\\1\end{pmatrix}$, we find
\begin{align}
 \left(g\mathscr B_\mu+\frac{g'}{2}B_\mu^Y\mathbf 1_2\right)\xi_0
 ={}&\frac12\Bigl[
 g(\mathscr B_\mu^1\sigma_1+\mathscr B_\mu^2\sigma_2)
 \nonumber\\
 &\quad+g\mathscr B_\mu^3\sigma_3+g'B_\mu^Y\mathbf 1_2
 \Bigr]\xi_0
 \nonumber\\
 ={}&\frac12
 \begin{pmatrix}
 g(\mathscr B_\mu^1-i\mathscr B_\mu^2)\\
 -g\mathscr B_\mu^3+g'B_\mu^Y
 \end{pmatrix}.
\end{align}
Thus the Higgs kinetic term is rewritten in terms of $\rho$, $\mathscr{B}_\mu=\mathscr{A}_\mu^h$, and $B_\mu^Y$ as
\begin{align}
 &(D_\mu[\mathscr A,B^Y]\Phi)^\dagger D^\mu[\mathscr A,B^Y]\Phi
 \nonumber\\
 &=\frac12(\partial_\mu\rho)^2 +\frac{\rho^2}{8}\Bigl[
 g^2\{(\mathscr B_\mu^1)^2+(\mathscr B_\mu^2)^2\}
+(g\mathscr B_\mu^3-g'B_\mu^Y)^2\Bigr].
 \label{kinetic1}
\end{align}
Therefore, if we introduce $W_\mu^\pm$, $Z_\mu$ and $A_\mu^{\rm em}$ as
\begin{align}
 \mathscr W_\mu^\pm&:=\mathscr{B}_\mu^\pm=\frac{1}{\sqrt2}(\mathscr{B}_\mu^1\mp i\mathscr{B}_\mu^2),
 \label{eq:ew-w}\\
 \mathscr Z_\mu&:=\cos\theta_W\,\mathscr{B}_\mu^3-
 \sin\theta_W\,B_\mu^Y,
 \label{eq:ew-z}\\
 A_\mu^{\rm em}&:=\sin\theta_W\,\mathscr{B}_\mu^3+
 \cos\theta_W\,B_\mu^Y,
 \label{eq:ew-photon}
\end{align}

Consequently, after the Weinberg rotation one finds the gauge transformation 
\begin{align}
 &\mathscr W_\mu^\pm \mapsto e^{\pm i\beta}\mathscr W_\mu^\pm,\qquad
 \mathscr Z_\mu\mapsto \mathscr Z_\mu,\nonumber\\
  &A_\mu^{\rm em}\mapsto A_\mu^{\rm em}+\frac1e\partial_\mu\beta.
\end{align}
This result shows that $\mathscr W^\pm$ carry charges $\pm1$, that $\mathscr Z$ and the Higgs particle $H$ are neutral, and that $A_\mu^{\rm em}$ is the electromagnetic gauge connection.
This makes the residual electromagnetic gauge structure explicit: the neutral $Z$ channel is locally gauge invariant, whereas $W^\pm$ remains charged and requires an electromagnetic dressing. Gauge/BRST-invariant composite operators for the $SU(2)\times U(1)$ Higgs theory and their renormalization have been studied directly in Refs.~\cite{DudalEtAl2023EW,Peruzzo2026EW}.

Comparing these equations with those in the usual unitary gauge, one sees that the transformation laws have the same form.
However, there is a decisive difference.
The ordinary $Z_\mu$ is a combination of fields in a chosen gauge. By contrast, $\mathscr Z_\mu$ is a composite vector field containing the Higgs field that is invariant under the original general $SU(2)_L\times U(1)_Y$ gauge transformations.
More specifically, $\mathscr Z_\mu$  can be written as
\begin{equation}\label{eq:old-11-3}
\mathscr Z_\mu=
\frac{4}{\sqrt{g^2+g'^2}} \frac{1}{\rho^2}
\operatorname{Im}(\Phi^\dagger D_\mu\Phi).
\end{equation}
The right-hand side is a gauge-invariant quantity constructed solely from the original fields. 
However, because the normalization contains $\rho^{-2}$, it is not defined at $\Phi=0$.

The gauge kinetic sector is
\begin{equation}
 \mathscr L_{\rm gauge}
 =-\frac12{\rm tr}\{\mathscr F_{\mu\nu}[\mathscr A]\mathscr F^{\mu\nu}[\mathscr A]\}
 -\frac14 B_{\mu\nu}^{Y}B_Y^{\mu\nu}.
\end{equation}
The Weinberg rotation is orthogonal in the neutral quadratic sector, so the canonically normalized neutral kinetic terms are equivalently expressed in the $Z_\mu$ and $A_\mu^{\rm em}$ basis, while the non-Abelian $SU(2)_L$ field strength retains the usual cubic and quartic interactions.

and
the Higgs kinetic term is rewritten in terms of $\rho$, $W_\mu^\pm$, $Z_\mu$, and  $A_\mu^{\rm em}$ as
\begin{align}
 (D_\mu\Phi)^\dagger D^\mu\Phi
 ={}&\frac12(\partial_\mu\rho)(\partial^\mu\rho)
 +\frac{\rho^2}{8}\Bigl[
 2g^2 W_\mu^+W^{-\mu}
 \nonumber\\
 &\qquad +(g^2+g'^2)Z_\mu Z^\mu
 +0\cdot A_\mu^{\rm em}A^{{\rm em}\mu}\Bigr].
\end{align}
where we have used the relationship between the Weinberg angle $\theta_W$ and the coupling constants $g,g'$:

\begin{align}
 \sin\theta_W&=\frac{g'}{\sqrt{g^2+g'^2}},
 &\cos\theta_W&=\frac{g}{\sqrt{g^2+g'^2}},\nonumber\\
 \tan\theta_W&=\frac{g'}{g}.&&
\end{align}
and hence
\begin{equation}
 e=g\sin\theta_W=g'\cos\theta_W=\frac{gg'}{\sqrt{g^2+g'^2}}.
\end{equation}

Therefore, at the radial minimum $\rho=v$, the masses of $W_\mu^\pm$, $Z_\mu$, and  $A_\mu^{\rm em}$ are respectively given by
\begin{equation}
 m_W=\frac{gv}{2},
 \qquad
 m_Z=\frac{\sqrt{g^2+g'^2}}{2}v,
 \qquad
 m_\gamma=0.
\end{equation}
The derivation can be carried out without choosing unitary gauge.
Neither non-vanishing VEV of the Higgs field $\langle\Phi\rangle$ nor gauge-symmetry breaking has been required at any stage of this derivation.

\subsection{The charged $W$ is not completely local gauge invariant}

On the other hand, $\mathscr W_\mu^+$  can be written as
\begin{equation}\label{eq:display-35}
\mathscr W_\mu^+= i \frac{2\sqrt2}{g\rho^2}\widetilde\Phi^\dagger D_\mu\Phi
\end{equation}
is invariant under $SU(2)_L$ but carries charge $+1$ with respect to electromagnetic $U(1)$.
Therefore, $\mathcal W_\mu^\pm$ are not fully local gauge-invariant quantities.

$W^\pm$ is charged under the residual electromagnetic $U(1)_{\rm em}$. Therefore, Gauss law prevents $W^\pm$ from being a strictly local gauge-invariant charged operator. 
A physical charged state requires an electromagnetic Dirac/Coulomb dressing, a Wilson-line dressing, or an appropriate asymptotic electromagnetic reference structure \cite{Dirac1955,Buchholz1986}.
 A thin Wilson line is sufficient as a formal gauge-invariant dressing, whereas a physically appropriate finite-energy/asymptotic charged state generally requires an infrared-compatible dressing, such as an appropriate Dirac/Coulomb or soft dressing. This is the electroweak example of the general partial-frame rule: Higgs-generated mass stiffness for a residual-$H$-charged coset mode does not by itself produce a local gauge-invariant interpolating operator for that charged mode.

On the other hand, the $Z$ and the radial Higgs operator $H$ are neutral and can therefore be constructed as local gauge-invariant operators. This distinction appears automatically in the QRF description.

\subsection{Making charged particles fully gauge invariant}

Here the electromagnetic Gauss law is indispensable.
$\mathscr W_\mu^+$ transforms as
\begin{equation}\label{eq:display-42}
\mathscr W_\mu^+(x)\longrightarrow e^{i\beta(x)}\mathscr W_\mu^+(x).
\end{equation}
Using the electromagnetic connection, construct, along a curve $C_{x_0x}$,
\begin{equation}\label{eq:display-43}
\mathbb W_\mu^+(x,x_0;C)
:=\mathscr W_\mu^+(x)
\exp\left[-ie\int_{x_0}^{x}A_\nu^{\rm em}(z)\,dz^\nu\right].
\end{equation}
Then
\begin{equation}\label{eq:display-44}
\mathbb W_\mu^+\longrightarrow
 e^{i\beta(x_0)}\mathbb W_\mu^+.
\end{equation}
If small gauge transformations at the reference point $x_0$ are restricted to the identity, this operator becomes gauge invariant.
However, if global or boundary transformations at $x_0$ are allowed, this phase remains. This is a characteristic feature of a physical state carrying charge.

Thus, the Higgs-QRF is a natural frame for removing the local gauge redundancy of $SU(2)_L$, but it does not automatically eliminate the electromagnetic Gauss law or boundary electric charges.

\subsection{Fermion mass generation can also be written without SSB in QRF language}

There is no need to restrict the Higgs-mechanism discussion to gauge bosons.
For a left-handed weak $SU(2)_L$ doublet $\psi_L=\begin{pmatrix}\nu_L\\e_L\end{pmatrix}$, define the relational doublet field $\Psi_L=\begin{pmatrix}N_L\\E_L\end{pmatrix}$ using the Higgs frame
\begin{equation}\label{eq:display-39}
\Psi_L(x)
=h(x)^\dagger\psi_L(x).
\end{equation}
The left-handed lepton doublet has hypercharge $Y_{\ell L}=-1/2$, whereas the Higgs doublet has $Y_{\Phi}=+1/2$. It transforms as
\begin{equation}\label{eq:display-5}
\psi_L(x) \mapsto
\Omega(x)e^{i\beta(x)Y}\psi_L(x), \quad Y_{\ell L}=-\frac12 .
\end{equation}
Then $\Psi_L(x)$ transforms as
\begin{equation}\label{eq:derivation-3}
\begin{aligned}
\Psi_L'=(h')^\dagger\psi_L'
=r_\beta h^\dagger \Omega^\dagger \Omega e^{i\beta Y}\psi_L
=e^{i\beta(T^3+Y)}\Psi_L.
\end{aligned}
\end{equation}
According to the Nakano--Nishijima--Gell-Mann rule the electric-charge operator $Q$ is equal to
\begin{equation}\label{eq:display-3}
Q=T^3+Y.
\end{equation}
Consequently, $\Psi_L(x)$ transforms as
\begin{equation}\label{eq:old-13-1}
\Psi_L(x)\mapsto e^{i\beta(x) Q}\Psi_L(x).
\end{equation}
The original $SU(2)_L$ doublet is converted into electric-charge eigenstates as seen from the Higgs reference frame.
The field $\Psi_L=h^\dagger \psi_L$ is invariant under local $SU(2)_L$ transformations, but it remains covariant under the residual right-acting $U(1)$ transformation and therefore should not be called completely gauge invariant before the electromagnetic dressing is specified.

For example, for leptons,
\begin{equation}\label{eq:display-41}
N_L\mapsto N_L,
\qquad E_L\mapsto e^{-i\beta}E_L.
\end{equation}
On a regular patch with $\rho\neq0$, $N_L$ is a neutral electroweak-invariant \emph{relational component}, while $E_L$ is $SU(2)_L$ invariant but carries electromagnetic charge.  The normalized representation through $h$ is nonpolynomial and becomes singular at $\rho=0$.  For quantum local-operator statements it is therefore preferable to use polynomial composites before dividing by $\rho$; explicitly,
\begin{equation}
 \widetilde\Phi^\dagger\psi_L=\frac{\rho}{\sqrt2}N_L,
 \qquad
 \Phi^\dagger\psi_L=\frac{\rho}{\sqrt2}E_L.
\end{equation}
The first operator is neutral and local, whereas the second still carries the residual electromagnetic charge and requires the corresponding physical dressing when used to create a charged state.

The right-handed electron transforms as
\begin{equation}\label{eq:display-6}
e_R(x)\mapsto
e^{-i\beta(x)}e_R(x).
\end{equation}
Similarly, the quark doublet is separated by the Higgs-QRF into relational fields carrying up-type and down-type charges. 
However, the $SU(3)_c$ gauge transformation property also remains for quarks.

For a fully relational rewriting, the Higgs fields and the left-handed fermion doublets must be expressed in the same Higgs frame.  Define
\begin{equation}
 \Phi^h:=h^\dagger\Phi=\frac{\rho}{\sqrt2}\xi_0,
 \qquad
 \widetilde\Phi^{\,h}:=h^\dagger\widetilde\Phi=\frac{\rho}{\sqrt2}\widetilde\xi_0,
\end{equation}
and, for the quark doublet $Q_L=\begin{pmatrix}u_L\\d_L\end{pmatrix}$, define the relational quark doublet $\mathcal Q_L=\begin{pmatrix}U_L\\D_L\end{pmatrix}$ by
\begin{equation}
 \mathcal Q_L
 :=h^\dagger Q_L,
\end{equation}

The Yukawa interactions are then
\begin{align}
 \mathscr L_Y={}&-y_e\,\overline\Psi_L\Phi^h e_R
 -y_d\,\overline{\mathcal Q}_L\Phi^h d_R
 -y_u\,\overline{\mathcal Q}_L\widetilde\Phi^{\,h}u_R
 +\mathrm{h.c.}\nonumber\\
 ={}&-\frac{\rho}{\sqrt2}
 \left(y_e\overline E_L e_R+y_d\overline D_L d_R+y_u\overline U_L u_R\right)
 +\mathrm{h.c.},
\end{align}
with the color indices in the quark terms contracted in the usual way.  Expanding $\rho=v+H$ gives
\begin{equation}
 m_f=\frac{y_f v}{\sqrt2}\qquad(f=u,d,e).
\end{equation}
Here $v$ is the classical/tree-level radial background (or fixed-length parameter), not the expectation value of a gauge-variant doublet.  Beyond tree level, neutral stable channels can be characterized by gauge-invariant spectral poles or gaps, while electromagnetically charged channels require the infrared-sensitive dressed spectral analysis described above.  The weak-isospin labels are relational with respect to the Higgs QRF, whereas the residual electromagnetic gauge transformation still has to be treated in charged sectors.

\section{Remarks on QRF-BEH mechanism}
\label{sec:remarks}

\subsection{Gauss law and the meaning of a ``physical gauge boson''}

In a Hamiltonian description, the Gauss constraint is schematically given by
\begin{equation}
 \mathcal G^a(x)
 =(\mathscr{D}_j E_j)^a(x)-\rho_{\rm matter}^a(x)\approx0.
\end{equation}
Physical states $|\mathrm{phys}\rangle$ satisfy the Gauss constraint
\begin{equation}
 \widehat{\mathcal G}^a(x)|\mathrm{phys}\rangle=0.
\end{equation}

This condition enforces invariance under the small/local gauge redundancies generated by the Gauss constraint. It does not exclude physical sectors carrying genuine asymptotic or global charges. In such charged sectors, the physical operator creating the state must include the corresponding nonlocal or asymptotic dressing, see e.g. \cite{KondoQRF2026}.
We therefore distinguish throughout between small gauge transformations generated by the local Gauss constraint, which are redundancies, and genuine global or asymptotic symmetry actions whose charges can label physical sectors.

A bare field carrying a local gauge index is not by itself a physical observable.
The Higgs frame converts such a gauge index into a relational label.
For a complete frame, the components of $\mathscr{B}_\mu$ are invariant under the original local gauge group;
for a partial frame, they transform only under the residual group that the Higgs field does not fix.

\textbf{The massive vector should therefore be understood as a gauge-invariant or residual-gauge-covariant relational excitation compatible with the Gauss constraint}, not as a gauge-variant field made physical by the spontaneous destruction of gauge redundancy.  This also explains why gauge-invariant composite operators are the natural objects from which the spectrum is extracted.

Therefore, a bare local color/gauge charge is not a local observable in the physical Hilbert space.
When a Higgs QRF is available, one can convert a gauge index into a QRF index and construct relational observables. In completely Higgsed $SU(2)$, a physical vector excitation should be regarded not as a particle carrying local gauge charge, but as a relational composite that is a gauge singlet while carrying a custodial/global index.
This is fully consistent with the FMS statement that the physical spectrum is described by gauge-invariant composites \cite{Maas2019,DudalEtAl2022}.
When a residual massless gauge group $H$ remains, its $H$ Gauss law also remains. Hence an $H$-charged state generally requires a nonlocal dressing or a boundary/asymptotic frame.

\subsection{Relational meaning of the would-be Nambu--Goldstone sector}

The degree-of-freedom reorganization has already been exhibited explicitly in Secs.~\ref{subsec:abelian-frame}--\ref{subsec:unitary-abelian}.
 In brief, the Higgs orientation supplies the frame coordinate entering the longitudinal sector of $\mathscr B_\mu$; unitary gauge is the representation $h=1$ in which the relational field is represented by the bare gauge potential. The distinct high-energy equivalence theorem is treated separately below as an amplitude statement rather than as another repetition of the ``eating'' picture.

\subsection{The equivalence theorem can also be re-read in QRF language}

The Nambu--Goldstone-boson equivalence theorem is a statement about properly normalized scattering amplitudes. Schematically, for $E\gg m_V$ and under the usual assumptions concerning gauge choice, renormalization, and external-state normalization,

\begin{equation}
 \mathcal M\!\left(V_L^{a_1}\cdots V_L^{a_n}\right)
 =C\,\mathcal M\!\left(\pi^{a_1}\cdots\pi^{a_n}\right)
 +O(m_V/E),
\end{equation}
where the finite factor $C$ depends on conventions and renormalization prescriptions. The theorem therefore does not follow merely from asserting that the derivative term $(i/g)h^{-1}\partial_\mu h$ ``dominates'' the longitudinal relational field.

The QRF interpretation is instead the following: the same high-energy longitudinal channel appearing in the equivalence theorem is naturally encoded in the frame-orientation sector entering the relational vector. This gives a relational interpretation of the theorem after the amplitude relation has been established, rather than a replacement for its field-theoretic derivation. A manifestly gauge-invariant formulation of the equivalence theorem in the Abelian Higgs model provides a useful comparison \cite{BoeykensDudalOosthuyse2025}.

\subsection{Global failure of the Higgs QRF: zeros, defects, and topology}

A limitation of the QRF construction appears when the Higgs frame becomes ill defined. For example, in an Abelian theory one writes
\begin{equation}
h(x)=\frac{\phi(x)}{|\phi(x)|},
\end{equation}
so at the spacetime point $x$ such that
\begin{equation}
\phi(x)=0
\end{equation}
the Higgs QRF cannot be defined.
A Nielsen--Olesen vortex core provides the standard example \cite{NielsenOlesen1973}.

In a non-Abelian theory as well, a global dressing $h[\Phi]$ may fail to exist at points where the orbit type of the Higgs field changes or on a topological defect.

\textbf{
In QRF language, a zero of the Higgs field is a locus at which the Higgs frame does not define an internal reference frame.
}
This is very important. When comparing the Higgs phase and confinement phase from a QRF viewpoint, rather than asking
\begin{equation}
\langle\Phi\rangle\neq0\ ?,
\end{equation}
a more physical criterion may be:
\textbf{how locally and globally a good relational gauge frame can be constructed}.

This criterion is not proposed as a universal local order parameter separating Higgs and confinement regions.  In Fradkin--Shenker-connected regimes it may vary smoothly without defining a sharp phase boundary.  Its role is instead to organize distinct gauge-invariant questions: local BEH mass generation, the long-distance availability and dressability of internal reference data, and global/topological obstructions to extending a chosen frame.

Equation~\eqref{eq:higgsframe} is not defined at $\rho=0$, so the normalized Higgs frame is a singular coordinate there.  A frame zero by itself, however, should not be identified automatically with a topologically protected defect: an isolated coordinate degeneration can be accidental or removable under a deformation.  A stable topological defect requires additional global information, such as a nontrivial homotopy class of the vacuum/orbit manifold, nontrivial bundle topology, or nontrivial patch-transition data.  In the Abelian gauge--Higgs theory a vortex core provides the standard example in which $\rho=0$ accompanies nontrivial winding of the phase frame.  Non-Abelian monopole-, vortex-, or texture-type sectors are likewise distinguished by the corresponding stabilizer, homotopy, and bundle data.  Thus QRF coordinate singularities and topological obstructions are related but conceptually distinct.

\textbf{The technical consequence is that local BEH mass generation can remain well defined in patches away from a defect, while line operators, topological sectors, confinement, and screening remain sensitive to the global patching of the frame.}

This separation makes it possible to discuss mass generation separately from global phase and topology.

\subsection{Global limitation of the unitary-gauge representation}

As already explained in Sec.~\ref{subsec:unitary-abelian}, the condition $h=1$ is a coordinate representation adapted to the Higgs QRF. The additional point needed here is global: where $\rho=0$ or the Higgs frame requires nontrivial patching, this representative cannot be imposed globally. The failure of unitary gauge therefore tracks the domain on which the Higgs-built dressing itself is defined.

\subsection{Relation to the Stueckelberg mechanism}
If only the angular part of a fixed-length Higgs field is retained, the mathematics is close to that of a Stueckelberg field. In the Abelian case,
$
 W_\mu=A_\mu-\frac1q\partial_\mu\vartheta
$
is precisely the Stueckelberg combination.
In the QRF interpretation, however, $h$ is not introduced as an ad hoc compensator but is regarded as a physical reference subsystem extracted from an already existing matter sector. This distinction is important.

Moreover, when non-Abelian massive Yang--Mills theory is treated as a Stueckelberg theory without a radial mode and taken as a fundamental four-dimensional QFT, high-energy problems of renormalizability/unitarity generally arise. QRF does not magically eliminate this problem. In a renormalizable Higgs theory, the radial mode and the complete gauge-invariant dynamics are important for the UV completion.

\subsection{Relation between QRF-BEH and the dressing field method}

In the dressing field method (DFM), a group-valued dressing $u$ is extracted from the field content and gauge-invariant composite fields such as
\begin{equation}
 \mathscr{A}^u=u^{-1}\mathscr{A}u+\frac{i}{g}u^{-1}du
\end{equation}
are constructed. Electroweak applications in which $u$ is constructed from a Higgs doublet have also long been explicit \cite{FournelEtAl2012}.

The mathematics of QRF and DFM is close, but QRF emphasizes the following physical meanings:
\begin{enumerate}[label=(\roman*)]
 \item the dressing is not merely an algebraic trick but an internal orientation standard;
 \item the frame itself is also a quantum system and fluctuates;
 \item a transformation between different frame choices becomes a physical relational change of perspective;
 \item the failure of a frame to exist globally is itself information about topology/confinement.
\end{enumerate}

The distinction between DFM and gauge fixing is discussed directly in Ref.~\cite{BerghoferFrancois2024}.
 The two operations should be separated explicitly. Constructing $\mathscr A_\mu^h=h^{-1}\mathscr A_\mu h+(i/g)h^{-1}\partial_\mu h$ from the original fields is a field-dependent construction of a gauge-invariant relational variable; it does not move the original field configuration onto a gauge slice. By contrast, choosing a representative for which $h=1$ and then writing $\mathscr A_\mu^h=\mathscr A_\mu$ is the unitary-gauge representation of that already defined relational variable. The algebra is closely related, but the operations have different meanings.

The matter-built construction is also close to, but not identical with, the field-space Higgs connection of Gomes, Hopfm\"uller and Riello \cite{Gomes:2018shn}. The present formulation emphasizes a group-valued dressing $h[\Phi]$ and the associated spacetime relational connection, whereas their construction emphasizes the field-space connection, horizontal symplectic geometry, and the treatment of charges. This distinction fixes the scope of the present QRF formulation without obscuring the close conceptual relation.

\section{Definition and dictionary for QRF-BEH}
\label{sec:definition}

\begin{widetext}
\begin{table*}[ht]
\caption{Dictionary between conventional and QRF/gauge-invariant descriptions of the BEH mechanism.}
\label{tab:qrf-dictionary}

\begin{ruledtabular}
\begin{tabular}{p{0.28\textwidth}p{0.66\textwidth}}
Conventional expression & QRF/gauge-invariant expression\\

$\langle\Phi\rangle\neq0$ & at tree level, a gauge-invariant radial background or fixed-length scale $v$\\
Gauge symmetry breaking & Higgs matter supplies an internal QRF along the relevant gauge-orbit directions\\
Vacuum chooses a direction & a component representation is chosen relative to a Higgs-QRF perspective\\
Would-be Nambu--Goldstone field & orientation coordinate of the Higgs QRF\\
Nambu--Goldstone is eaten & connection and frame orientation are reorganized into a relational vector\\
Longitudinal vector mode & includes the frame-orientation derivative sector\\
Massive gauge boson& stable local gauge-invariant spin-one pole/gap in complete or neutral channels; resonance data for unstable states; infrared-sensitive dressed spectral structure in residual long-range charged channels\\
$G\to H$ & the Higgs data select the orbit/coset subspace while residual $H$ basis freedom remains\\
Unitary gauge & representation adapted to the Higgs QRF, $h=1$\\
Higgs particle & gauge-invariant radial excitation\\
Higgs defect & locus or patching obstruction where the Higgs QRF degenerates or fails globally\\
Higgs/confinement comparison & local mass generation, long-distance frame properties, and global/topological diagnostics are treated separately\\
\end{tabular}
\end{ruledtabular}
\end{table*}
\end{widetext}

The preceding constructions motivate the following operational definition. In a gauge theory with matter, the BEH mechanism is realized when matter supplies internal reference data along at least part of the gauge orbit and the matter kinetic term generates a positive-semidefinite mass-squared quadratic form for the corresponding components of the relational connection. At tree level this quadratic form reproduces the usual BEH mass matrix. Quantum mechanically, a stable complete-frame or residual-neutral one-particle state may be identified through an isolated pole or spectral gap of an appropriate renormalized local gauge-invariant composite correlator, while an unstable state is characterized by resonance data such as a complex pole. A residual-$H$-charged excitation requires an appropriately dressed, generally nonlocal or asymptotic operator; if the residual gauge interaction is long ranged, its spectral measure need not contain an isolated one-particle pole. The Higgs-generated mass-squared form, the existence/locality of a physical interpolating operator, and the detailed quantum spectral singularity are therefore distinct statements.

This definition does not require a gauge-variant expectation value $\langle\Phi\rangle\neq0$ or the physical spontaneous breaking of local gauge redundancy. Genuine global, custodial, or asymptotic symmetries remain separate questions.
See Table.~\ref{tab:qrf-dictionary}.

\section{Conclusion and discussion}
\label{sec:conclusion}

\textbf{The BEH mechanism can be stated without assigning physical spontaneous breaking to a local gauge redundancy.} 
Higgs matter supplies a dynamical internal reference frame, and the gauge connection relative to that frame is the natural relational variable. 

At the classical level, \textbf{the Higgs kinetic term induces a positive-semidefinite mass-squared quadratic form on the appropriate gauge-orbit components of the relative connection and reproduces the standard tree-level BEH masses.}

neutral vector excitation may be characterized by a local gauge-invariant composite pole or spectral gap.  A residual-charged excitation instead requires an appropriately dressed spectral object; when the residual gauge field is massless, the relevant singularity need not be an isolated one-particle pole and can have infraparticle-type threshold structure.

\textbf{Would-be Nambu--Goldstone fields are thereby interpreted as coordinates of the Higgs-frame orientation rather than as physical massless particles that subsequently disappear.} 
In a complete frame, these orientation variables enter the longitudinal sector of the gauge-invariant relational vector; in a partial frame, a residual gauge group remains. 
This formulation is consistent with Elitzur's theorem, gauge-invariant BEH variables, the FMS composite spectrum, and the residual electromagnetic Gauss law.

The conclusions therefore mirror the three structural statements announced in the Introduction: 

(i) complete frames give a local relational description on regular patches, 

(ii) partial frames retain a stabilizer-sector gauge structure whose algebra is the kernel of the Higgs-generated mass matrix, and 

(iii) the electroweak application leaves the electromagnetic Gauss law explicit rather than converting charged modes into strictly local observables.

The QRF language also separates local mass generation from global questions. Zeros of the Higgs field, nontrivial patching, line operators, generalized symmetries, and the long-distance availability of reference frames probe information that is not contained in the local BEH mass term.

The \textbf{novelty of this paper} is not the bare statement that local gauge symmetry need not be physically broken. Rather, this paper provides a unified relational organization in which 

(i) complete and partial Higgs dressings are treated as internal reference structures, 

(ii) the BEH mass matrix is identified geometrically with the metric induced on the gauge orbit, 

(iii) FMS composite interpolators and Kondo-type massive modes  are related directly to relational connection components, and 

(iv) the electroweak residual Gauss law is retained explicitly, thereby separating Higgs-generated stiffness from the existence of local physical charged operators.

The exact local dressing identities and FMS leading-pole correspondence used here overlap with established gauge-invariant descriptions \cite{FMS1980,FMS1981,FournelEtAl2012,DudalEtAl2022}; in particular, identifying the Higgs dressing as reference data does not by itself constitute a new quantum prediction. The present paper does not claim a new numerical quantum-spectrum prediction beyond the established Higgs/FMS framework.
Its concrete contribution is the unified complete/partial-frame organization, frame covariance, and the explicit residual-Gauss-law treatment.  Quantitative QRF-specific tests---for example correlations of two independent Higgs frames, frame-dependent operator mixing with frame-independent spectral data, or a lattice measurement of the coherence diagnostics below---are left as targeted future work.



In the QRF approach the following gauge-invariant quantities should be used instead of the SSB order parameter $\langle\Phi\rangle$.

First, the radial sector:
\begin{equation}
 C_H(x-y)
 =\left\langle
 \left(\Phi^\dagger\Phi(x)-\langle\Phi^\dagger\Phi\rangle\right)
 \left(\Phi^\dagger\Phi(y)-\langle\Phi^\dagger\Phi\rangle\right)
 \right\rangle.
\end{equation}
The scalar pole/mass gap is read from this.

Second, the relational vector correlator:
\begin{equation}
 C_{\mu\nu}^{ab}(x-y)
 =\langle \mathcal O_\mu^a(x)\mathcal O_\nu^b(y)\rangle,
 \qquad
 \mathcal O_\mu^a\propto\rho^2\mathscr{B}_\mu^a.
\end{equation}

For a complete frame or residual-neutral channel, a stable physical vector pole or spectral gap can be read from such a local gauge-invariant correlator.  A residual-$H$-charged channel instead requires an appropriately dressed spectral observable, and in the presence of long-range residual gauge dynamics its spectral measure need not possess an isolated one-particle pole.

Third, for a complete frame, define the following semi-local \emph{QRF coherence diagnostic}, which is a Higgs-endpoint/open-Wilson-line observable rather than a new independent class of gauge-invariant operator:
\begin{equation}
 C_{\rm QRF}(x,y;\gamma)
 =\left\langle
 \frac1{d_R}\mathrm{Re}\,\mathrm{tr}_R
 \left[h^\dagger(x)U_\gamma[\mathscr A](x\leftarrow y)h(y)\right]
 \right\rangle,
 \label{eq:qrfcoherence}
\end{equation}
Here $h$ and the transporter must be taken in the same representation $R$ (in the fundamental $SU(2)$ example, $R=\mathbf2$), and $d_R:=\dim R$. The trace also specifies the custodial-index contraction; a gauge-invariant endpoint operator does not automatically furnish a path-independent coherence observable.

For the complete-frame relational connection $\mathscr B_\mu=\mathscr A_\mu^h$, the corresponding parallel transporter obeys the exact identity
\begin{equation}
 U_\gamma[\mathscr B](x\leftarrow y)
 =h^\dagger(x)U_\gamma[\mathscr A](x\leftarrow y)h(y),
\end{equation}
and hence
\begin{equation}
 C_{\rm QRF}(x,y;\gamma)
 =\left\langle\frac1{d_R}\mathrm{Re}\,\mathrm{tr}_R
 U_\gamma[\mathscr B](x\leftarrow y)\right\rangle.
\end{equation}
This makes explicit that the proposed diagnostic is the Wilson line of the relational connection.  Its value is generally path dependent, and its continuum definition requires the usual treatment of Wilson-line self-energy/perimeter divergences; no path-independent universal correlation length is assumed.  For a partial frame, the lift $h\sim hr$ has a residual $H$ ambiguity and the same endpoint expression is not generally invariant under independent $r(x)$ and $r(y)$. One should instead use an observable constructed directly from the coset variable. For an adjoint $SU(2)$ frame a natural example is
\begin{equation}
 C_n(x,y;\gamma)
 =\left\langle 2\,\mathrm{tr}\!\left[
 \bm n(x)U_\gamma(x\leftarrow y)
 \bm n(y)U_\gamma(y\leftarrow x)
 \right]\right\rangle.
 \label{eq:partialqrfcoherence}
\end{equation}
This definition does not require a choice of local lift $h$. Both complete- and partial-frame correlators remain path dependent in general, and their continuum Wilson-line factors require the usual treatment of self-energy/perimeter divergences. Any coherence length extracted from them should therefore be regarded as a proposed, prescription-dependent diagnostic until its robustness under path choices and renormalization prescriptions has been demonstrated.

Fourth, defect density, frame-zero probability, holonomy distributions, and line endability can be monitored as complementary diagnostics. Bare defect densities or frame-zero probabilities may be cutoff dependent in the continuum limit, so physically robust versions should be tied to renormalized densities, percolation observables, homotopy/bundle classes, or line-endability data.

The important point is not to force all of these into a single scalar order parameter. 
When Fradkin--Shenker continuity \cite{FradkinShenker1979} permits it, Higgs-like and confinement-like regimes can be continuously connected in a crossover-like manner.
Concrete QRF-specific tests that go beyond the present unification include comparing two independent matter-built frames through the gauge-invariant relative orientation $s=h_1^{-1}h_2$, exhibiting frame-dependent operator mixing with invariant spectral energies in a solvable example, and measuring the renormalized lattice coherence diagnostic with controlled path/perimeter subtraction.
The connection between confinement phase and the Higgs phase will be discussed in more detail in a subsequent paper.

\section*{Acknowledgments}
This work was supported by Grant-in-Aid for Scientific Research, JSPS KAKENHI Grant
Number (C) No.23K03406.
During preparation of this manuscript, the author used ChatGPT (OpenAI; GPT-5.6 Sol and GPT-5.5) for language editing, restructuring selected passages, and preliminary consistency checks of derivations and bibliography.
The author independently rederived or verified all scientific suggestions against the relevant calculations and cited sources, reviewed all AI-assisted text, and takes full responsibility for the accuracy, citations, scientific claims, and conclusions of the manuscript.

\section*{Data Availability}
No data were created or analyzed in this study.

\appendix
\section{Derivation of the adjoint--Higgs model}
\label{app:adj-Higgs}

The Lie-algebra $su(2)$-valued connection is decomposed as
\begin{align}
 &\mathscr B_\mu
 = \mathscr B_\mu^\parallel+\mathscr B_\mu^\perp ,
 &\mathscr B_\mu^\parallel=B_\mu^3T_3, \quad
 \mathscr B_\mu^\perp=B_\mu^1T_1+B_\mu^2T_2, 
\end{align}
for its stabilizer and coset parts.
Then $\mathscr B_\mu^\parallel$ is the residual $U(1)$ connection, whereas $\mathscr B_\mu^\perp$ is a coset field transforming homogeneously under the residual $U(1)$. Hence the adjoint Higgs field realizes the partial relationalization
\begin{equation}
 SU(2)
 \quad\longrightarrow\quad
 \underbrace{SU(2)/U(1)}_{\text{frame supplied}}
 +
 \underbrace{U(1)}_{\text{unfixed residual frame}}
 \label{eq:partialrelationalization-appendix}
\end{equation}
displayed above.

Since 
\begin{equation}
\bm n=hT_3h^\dagger,
\end{equation} 
moving to the frame $h$ gives
\begin{equation}
 h^\dagger \mathscr{D}_\mu[\mathscr A]\bm n\,h
 =\mathscr{D}_\mu[h^\dagger\mathscr A h](h^\dagger\bm n\,h)
 =\mathscr{D}_\mu[\mathscr B]T_3,
 \label{eq:covtransformn}
\end{equation}
for the covariant derivative of the frame direction.

Because $T_3$ is constant,
\begin{equation}
 \mathscr{D}_\mu[\mathscr B]T_3
 =
 -ig[\mathscr B_\mu,T_3]
 = gB_\mu^2T_1-gB_\mu^1T_2
 \label{eq:DBT3}
\end{equation}
holds.
Moreover,
\begin{equation}
 [\mathscr B_\mu^\parallel,T_3]=0,
 \label{eq:parallelcommute}
\end{equation}
so that
\begin{equation}
 \mathscr{D}_\mu[\mathscr B]T_3
 =\partial_\mu T_3 -ig[\mathscr B_\mu^\perp,T_3]
 =-ig[\mathscr B_\mu^\perp,T_3],
 \label{eq:DBT3perp}
\end{equation}
only the coset component contributes.
Because for $\bm\phi=v\bm n$
\begin{equation}
 \mathscr{D}_\mu[\mathscr A]\bm\phi
 =
 v\mathscr{D}_\mu[\mathscr A]\bm n,
 \label{eq:DphiDn}
\end{equation}
the Higgs kinetic term becomes
\begin{equation}
 \frac12\mathscr{D}_\mu[\mathscr A]\bm\phi\cdot \mathscr{D}^\mu[\mathscr A]\bm\phi
 =
 \frac{v^2}{2}
 \mathscr{D}_\mu[\mathscr B]T_3\cdot
 \mathscr{D}^\mu[\mathscr B]T_3,
 \label{eq:kineticframe}
\end{equation}
in the QRF frame.
Using (\ref{eq:DBT3}), 
one obtains\begin{equation}
 \boxed{
 \frac12\mathscr{D}_\mu\bm\phi\cdot \mathscr{D}^\mu\bm\phi
 =
 \frac12(gv)^2
 \left[
 B_\mu^1B^{\mu1}
 +
 B_\mu^2B^{\mu2}
 \right]
 },
 \label{eq:massfromkinetic}
\end{equation}
the explicit mass term.
Hence
\begin{equation}
 M_W=gv,
 \label{eq:MW}
\end{equation}
whereas $B_\mu^3$ receives no Higgs-generated mass term.
Equivalently, the mass matrix is
\begin{equation}
 (M^2)_{AB}
 =
 g^2v^2
 \left(
 \delta_{AB}-n_An_B
 \right)
,
 \label{eq:massmatrixadjoint}
\end{equation}
which is precisely the projector onto the tangent space of $S^2=SU(2)/U(1)$.
From the QRF viewpoint, the correct statement is not that a ``broken generator becomes massive, but that a relative-motion stiffness is generated along those gauge-orbit directions for which the Higgs QRF supplies reference data.

The paper \cite{Kondo2016PLB} introduces
\begin{equation}
 \mathscr W_\mu
 =
 -\frac{i}{g}
 [\bm n,\mathscr{D}_\mu[\mathscr A]\bm n]
 \label{eq:KondoW}
\end{equation}
as the massive mode.
Transforming this field to the QRF frame gives
\begin{equation}
 h^\dagger\mathscr W_\mu h
 =
 -\frac{i}{g}
 [T_3,\mathscr{D}_\mu[\mathscr B]T_3]
 \label{eq:Wframe1}
\end{equation}
directly.
Using Eq.~\eqref{eq:DBT3}, this becomes
\begin{equation}
 h^\dagger\mathscr W_\mu h
 =
 -[T_3,[\mathscr B_\mu,T_3]]
 \label{eq:Wframe2}
\end{equation}
For $SU(2)$, the double commutator is the projector onto the components orthogonal to $T_3$, so
\begin{equation}
 [T_3,[T_3,\mathscr B_\mu]]
 =
 \mathscr B_\mu^\perp
 \label{eq:doublecommproj}
\end{equation}
holds. Therefore
\begin{equation}
 h^\dagger\mathscr W_\mu h
 =
 \mathscr B_\mu^\perp
 \label{eq:W=Bperp-appendix}
\end{equation}
follows.
Thus, in QRF language, the massive mode of Ref.~\cite{Kondo2016PLB} is
\begin{equation}
\begin{aligned}
 \mathscr W_\mu={}&\text{coset-relative part of the connection}\\
 &\text{with respect to the partial Higgs QRF}
 \end{aligned}
 \label{eq:Winterpretation}
\end{equation}
and nothing more is required to identify it.
Consequently, the mass identity follows immediately:
\begin{equation}
 \frac12
 \mathscr{D}_\mu[\mathscr A]\bm\phi\cdot
 \mathscr{D}^\mu[\mathscr A]\bm\phi
 =
 \frac12(gv)^2
 \mathscr W_\mu\cdot\mathscr W^\mu
 \label{eq:massidentityW}
\end{equation}
which is exactly the fixed-length identity of the adjoint-Higgs construction \cite{Kondo2016PLB}.


From the definition,
\begin{equation}
 \mathscr W_\mu\cdot\bm n
 =
 -\frac{i}{g}
 [\bm n,\mathscr{D}_\mu\bm n]\cdot\bm n
 =0
 \label{eq:Worthogonalcalc}
\end{equation}
and hence
\begin{equation}
 \bm n\cdot\mathscr W_\mu=0
 \label{eq:Worthogonal}
\end{equation}
follows.
From the QRF viewpoint, $\bm n$ specifies the residual $U(1)$ direction, whereas $\mathscr W_\mu$ lies in the orthogonal tangent direction
\begin{equation}
 \mathscr W_\mu\in T_{\bm n}(SU(2)/U(1))
 \label{eq:Wtangent}
\end{equation}
This also makes it natural that $\mathscr W_\mu$ has two independent internal components.

\section{Derivation of Eq.~(\ref{kinetic1}).}
\label{app:ew-derivation}

The derivation can be carried out without choosing unitary gauge.

We use the conventions already adopted in Sec.~\ref{subsec:ew-frame},

\begin{align}
 \Phi&=\frac{\rho}{\sqrt2}h\xi_0, \quad
 \xi_0=\begin{pmatrix}0\\1\end{pmatrix},
 \\
 \mathscr{B}_\mu
 &=h^\dagger\mathscr{A}_\mu h
   +\frac{i}{g}h^\dagger\partial_\mu h, \quad
 \mathscr{B}_\mu =\mathscr{B}_\mu^a\frac{\sigma_a}{2},
\end{align}
and the electroweak covariant derivative acting on the Higgs doublet of hypercharge $Y=1/2$,
\begin{align}
 D_\mu\Phi
 =\left(
 \partial_\mu-ig\mathscr{A}_\mu
 -i\frac{g'}{2}B_\mu^Y
 \right)\Phi.
\end{align}

First, the definition of the relational connection $\mathscr{B}_\mu$ gives
\begin{align}
 -ig\mathscr{B}_\mu
 =-ig h^\dagger\mathscr{A}_\mu h
   +h^\dagger\partial_\mu h
=h^\dagger
   \left(\partial_\mu-ig\mathscr{A}_\mu\right)h,
\end{align}
which leads to
\begin{align}
 \left(\partial_\mu-ig\mathscr{A}_\mu\right)h
 =-ig\,h\mathscr{B}_\mu.
\end{align}

Now differentiate $\Phi=(\rho/\sqrt2)h\xi_0$ explicitly:
\begin{align}
 D_\mu\Phi
 ={}&\frac{1}{\sqrt2}\Bigl[
 (\partial_\mu\rho)h\xi_0
 +\rho(\partial_\mu-ig\mathscr A_\mu)h\xi_0
-i\rho\frac{g'}{2}B_\mu^Yh\xi_0\Bigr]
 \nonumber\\
 ={}&\frac{h}{\sqrt2}\Bigl[
 (\partial_\mu\rho)\xi_0
 -i\rho\left(g\mathscr B_\mu+\frac{g'}{2}B_\mu^Y\mathbf 1_2\right)\xi_0
 \Bigr].
\end{align}
It is convenient to set, only within this derivation,
\begin{align}
 K_\mu
 :=g\mathscr{B}_\mu+\frac{g'}{2}B_\mu^Y\mathbf{1}_2.
\end{align}
Since $h^\dagger h=\mathbf{1}_2$, $\xi_0^\dagger\xi_0=1$, and $K_\mu^\dagger=K_\mu$, one obtains
\begin{align}
 (D_\mu\Phi)^\dagger D^\mu\Phi
 &=\frac12
 \left[
 (\partial_\mu\rho)\xi_0^\dagger
 +i\rho\xi_0^\dagger K_\mu
 \right]
 \left[
 (\partial^\mu\rho)\xi_0
 -i\rho K^\mu\xi_0
 \right]
 \nonumber\\
 &=\frac12(\partial_\mu\rho)(\partial^\mu\rho)
 +\frac{\rho^2}{2}\,\xi_0^\dagger K_\mu K^\mu\xi_0.
\end{align}
The two mixed terms cancel exactly:
\begin{align}
 &-\frac{i\rho}{2}(\partial_\mu\rho)
 \xi_0^\dagger K^\mu\xi_0
 +\frac{i\rho}{2}(\partial^\mu\rho)
 \xi_0^\dagger K_\mu\xi_0
 =0.
\end{align}
Thus the remaining task is to evaluate $K_\mu\xi_0$.

Using
\begin{align}
 \sigma_1\xi_0&=\begin{pmatrix}1\\0\end{pmatrix},
 &
 \sigma_2\xi_0&=\begin{pmatrix}-i\\0\end{pmatrix},
 &
 \sigma_3\xi_0&=-\xi_0,
\end{align}
we find
\begin{align}
 K_\mu\xi_0
 &=\frac12
 \left[
 g\mathscr{B}_\mu^1\sigma_1
 +g\mathscr{B}_\mu^2\sigma_2
 +g\mathscr{B}_\mu^3\sigma_3
 +g'B_\mu^Y\mathbf{1}_2
 \right]\xi_0
 \nonumber\\
 &=\frac12
 \begin{pmatrix}
 g\bigl(\mathscr{B}_\mu^1-i\mathscr{B}_\mu^2\bigr)
 \\
 -g\mathscr{B}_\mu^3+g'B_\mu^Y
 \end{pmatrix}.
\end{align}
Therefore
\begin{align}
 \xi_0^\dagger K_\mu K^\mu\xi_0
 &=(K_\mu\xi_0)^\dagger(K^\mu\xi_0)
 \nonumber\\
 &=\frac14\Bigl[
 g^2(\mathscr B_\mu^1+i\mathscr B_\mu^2)
 (\mathscr B^{1\mu}-i\mathscr B^{2\mu})
 \nonumber\\
 &\quad+(-g\mathscr B_\mu^3+g'B_\mu^Y)
 (-g\mathscr B^{3\mu}+g'B^{Y\mu})\Bigr]
 \nonumber\\
 &=\frac14\Bigl[
 g^2\mathscr B_\mu^1\mathscr B^{1\mu}
 +g^2\mathscr B_\mu^2\mathscr B^{2\mu}
 \nonumber\\
 &\quad +(g\mathscr B_\mu^3-g'B_\mu^Y)
 (g\mathscr B^{3\mu}-g'B^{Y\mu})\Bigr].
\end{align}
In the last step the cross terms between the $1$ and $2$ components cancel.
Substituting this expression back gives
\begin{align}
 (D_\mu\Phi)^\dagger D^\mu\Phi
 ={}&\frac12(\partial_\mu\rho)(\partial^\mu\rho)
 \nonumber\\
 &+\frac{\rho^2}{8}\Bigl[
 g^2\mathscr B_\mu^1\mathscr B^{1\mu}
 +g^2\mathscr B_\mu^2\mathscr B^{2\mu}
 \nonumber\\
 &\qquad +(g\mathscr B_\mu^3-g'B_\mu^Y)
 (g\mathscr B^{3\mu}-g'B^{Y\mu})\Bigr].
\end{align}
Equivalently, with the shorthand $(\mathscr{B}_\mu^a)^2:=\mathscr{B}_\mu^a\mathscr{B}^{a\mu}$ and $(B_\mu^Y)^2:=B_\mu^Y B^{Y\mu}$, this is precisely Eq.~(\ref{kinetic1}).

\section{Local Jacobian and notation conventions}
\label{app:jacobian}

For an $SU(2)$ fundamental doublet, the four real scalar components can locally be decomposed into one radial coordinate $\rho$ and three angular coordinates on $S^3\simeq SU(2)$. Accordingly, away from $\rho=0$, the scalar-field measure contains the local polar-coordinate Jacobian
\begin{equation}
 \mathcal D\Phi\,\mathcal D\Phi^\dagger
 \propto \mathcal D h\,\mathcal D\rho\,\prod_x\rho(x)^3,
\end{equation}
up to normalization and regularization-dependent factors.  On a regular complete-frame patch, quotienting the local $SU(2)$ orbit volume leaves the formal radial factor $J[\rho]\propto\prod_x\rho(x)^3$.  This statement is local and regulator dependent; it does not by itself define a globally valid renormalized path-integral measure.  The $\rho=0$ strata and nontrivial global patching lie outside this coordinate chart.


\section{What it means for the QRF to be a quantum system}
\label{sec:quantum-frame}

In the classical discussion of a dressing field, $h[\Phi]$ can be treated simply as a function of the fields.  In quantum theory the Higgs degrees of freedom fluctuate, so the frame is dynamical and quantum; this fact alone, however, is not the full QRF content.  The QRF interpretation also uses the nontrivial gauge transformation of the reference degrees of freedom, the construction of relational observables relative to them, and the admissible frame transformations described in Sec.~\ref{sec:qrf-dictionary}.  Frame quality is then an additional quantum property rather than a consequence of fluctuation alone.

Therefore, the physical vector excitation is not $\mathscr A_\mu$ relative to an externally fixed internal direction, but a relational observable $\mathscr A_\mu^h$ relative to a quantum Higgs frame $\mathscr h$. Conceptually, one makes the division
\begin{equation}
 \begin{aligned}
 \text{system}&=\text{gauge connection},\\
 \text{QRF}&=\text{Higgs orientation}.
 \end{aligned}
\end{equation}
and measures their relative quantity.

This split is only a kinematical organization of field degrees of freedom used to construct relational observables.  It does not assume that the Gauss-constrained physical Hilbert space factorizes as a tensor product of an independent gauge-field subsystem and an independent Higgs-frame subsystem; such a factorization is generally obstructed in gauge theory and depends on boundary/edge-mode choices.

From this standpoint, \textbf{the BEH mechanism may be interpreted as a relational description in which the dynamical Higgs orientation supplies internal frame data.}
In continuum QFT, however, a normalized nonlinear object such as $h[\widehat\Phi]\sim\widehat\Phi/(\widehat\Phi^\dagger\widehat\Phi)^{1/2}$ should be regarded as a formal relational coordinate unless a regulator and composite-operator prescription are specified. Pointwise products, inverse square roots, the $\rho=0$ strata, and renormalization all require care. 
A simple regulated realization is available on the lattice.  Let $h_x$ be the complete Higgs frame constructed from the site Higgs variable and let $U_{x,\mu}$ be a link variable.  Define
\begin{equation}
 U^h_{x,\mu}:=h_x^\dagger U_{x,\mu}h_{x+\hat\mu}.
\end{equation}
Under $h_x\mapsto\Omega_x h_x$ and $U_{x,\mu}\mapsto\Omega_xU_{x,\mu}\Omega_{x+\hat\mu}^\dagger$, one has $U^h_{x,\mu}\mapsto U^h_{x,\mu}$.  Thus, at finite regulator, the dynamical Higgs degree of freedom supplies a genuine quantum internal frame relative to which the gauge transporter is measured, without first requiring a pointlike renormalized operator $h[\widehat\Phi]$ in the continuum.

 For quantum spectral statements it is therefore safer to place polynomial or otherwise renormalized gauge-invariant composites in the foreground. For example, in the $SU(2)$ fundamental model the vector interpolator $\mathcal O_\mu^a=(g/2)\rho^2\mathscr B_\mu^a$ is a suitable local composite, whereas $h$ and $\mathscr B_\mu$ are most naturally viewed as relational patch variables. 
The all-order renormalization and Ward-identity analysis of Higgs-model composite operators provides the relevant technical setting \cite{DudalEtAl2022}; manifestly gauge-invariant quantum-field constructions offer complementary renormalization perspectives \cite{Quadri2024}. 
Accordingly, the statement that the vector mass term arises from the classical Higgs-orbit metric is logically prior to, and distinct from, the quantum determination of the physical spectrum from renormalized composite correlators.  The gauge bosons are not massive merely because the reference frame is quantum.  Existing Ward identities and composite-operator renormalization results should be used when identifying the propagating spin-one channel and separating it from longitudinal or constraint contributions \cite{DudalEtAl2022,Peruzzo2026EW}.

The fact that the Higgs field is quantum also does not by itself guarantee a high-quality QRF.  We \emph{propose} that the quality of a semiclassically well-resolved Higgs frame be quantified by suitable renormalized measures of orientation uncertainty, relational coherence, or group asymmetry, and compared with the control parameters of the FMS expansion \cite{deLaHametteLudescherMuller2022}.  Constructing a regulator-independent continuum-QFT frame-quality measure is left for future work.

\bibliography{ep283v1}

\end{document}